\documentclass[longauth]{aa}  

\usepackage{graphicx}
\usepackage{multirow}
\usepackage{xcolor}
\usepackage[caption=false]{subfig}
\usepackage{bm}

\usepackage{txfonts}

\usepackage[colorlinks=true, allcolors=blue]{hyperref}

\usepackage{lscape}

\usepackage{soul}

\usepackage{amssymb}
\usepackage{pifont}

\begin{document} 

    \title{EWOCS-IX: JWST/NIRCam observations of Westerlund 2 - Identification of candidate substellar members
    }
    \titlerunning{EWOCS-VI: JWST/NIRCam observations of Westerlund 2}
    \authorrunning{Almendros-Abad et al.}
    
   \author{ V. Almendros-Abad 
   \inst{1}\thanks{Corresponding author: \texttt{victor.almendrosabad@inaf.it}}, M.G. Guarcello \inst{1}, K. Mu\v{z}i\'c\inst{2}, A. Scholz \inst{3}, M. Andersen \inst{4}, A. Bayo \inst{4}, W. Best \inst{5}, D. Capela \inst{3}, M. Gennaro \inst{6}, A. Ginsburg \inst{7}, J.B. Lovell \inst{8}, K. Monsch \inst{8}, E. Moraux \inst{9}, L. Prisinzano \inst{1}, T. Rom \inst{10,9}, E. Sabbi \inst{11,12}, P. Zeidler \inst{13}, C. Argiroffi \inst{14,1}, R. Bonito \inst{1}, D. Calzetti \inst{15}, V. Cusimano \inst{14,1}, F. Damiani \inst{1}, J.J. Drake \inst{16}, T.J. Haworth \inst{17}, N.D. Richardson \inst{18}, M. Robberto \inst{7,19}, S. Sciortino \inst{1}, N.J. Wright \inst{20}, T. Ziliotto \inst{21} 
          }

        \institute{Istituto Nazionale di Astrofisica (INAF) - Osservatorio Astronomico di Palermo, Piazza del Parlamento 1, 90134 Palermo, Italy
        \and
        Instituto de Astrofísica e Ciências do Espaço, Faculdade de Ciências, Universidade de Lisboa, Ed. C8, Campo Grande, 1749-016 Lisbon, Portugal
        \and
        SUPA, School of Physics \& Astronomy, University of St Andrews, North Haugh, St Andrews KY16 9SS, UK
        \and
        European Southern Observatory, Karl-Schwarzschild-Strasse 2, 85748 Garching bei München, Germany
        \and
        The University of Texas at Austin, Department of Astronomy, 2515 Speedway, C1400, Austin, TX 78712, USA
        \and
        Space Telescope Science Institute, 3700 San Martin Dr, Baltimore, MD 21218, USA
        \and
        Department of Astronomy, University of Florida, P.O. Box 112055, Gainesville, FL 32601, USA
        \and
        Center for Astrophysics $\vert$ Harvard \& Smithsonian, 60 Garden Street, Cambridge, MA 02138, USA
        \and
        Univ. Grenoble Alpes, CNRS, IPAG, 38000 Grenoble, France
        \and
        University of Split, Faculty of Science, Department of Physics, Ru\dj{}era Bo\v{s}kovi\'{c}a 33, 21000 Split, Croatia
        \and
        Gemini Observatory/NSF NOIRLab, 950 N. Cherry Ave., Tucson, AZ 85719, USA
        \and
        Steward Observatory, University of Arizona, 933 North Cherry Avenue, Tucson, AZ 85721, USA
        \and
        AURA for the European Space Agency (ESA), ESA Office, Space Telescope Science Institute, 3700 San Martin Drive, Baltimore, MD 21218, USA
        \and
        Department of Physics and Chemistry, University of Palermo, Piazza del Parlamento 1, 90134 Palermo, Italy
        \and
        Department of Astronomy, University of Massachusetts Amherst, 710 North Pleasant Street, Amherst, MA 01003, USA
        \and
        Lockheed Martin Solar and Astrophysics Laboratory, 3251 Hanover Street, Palo Alto, CA 94304, USA
        \and
        Astronomy Unit, School of Physics and Astronomy, Queen Mary University of London, London E1 4NS, UK
        \and
        Department of Physics and Astronomy, Embry-Riddle Aeronautical University, 3700 Willow Creek Rd, Prescott, AZ 86301, USA
        \and
        Johns Hopkins University, 3400 N. Charles St., Baltimore, MD 21218, USA
        \and
        Astrophysics Research Centre, Keele University, Keele ST5 5BG, UK
        \and
        NSF’s National Optical-Infrared Astronomy Research Laboratory, 950 N. Cherry Ave., Tucson, AZ 85719, USA
        }
        
   \date{Received; accepted}

  \abstract
   {Investigating substellar populations in young massive clusters offers crucial insights into the formation of brown dwarfs (BDs) and the role of environmental conditions in shaping their properties. Westerlund~2 (Wd2), as one of the nearest dense and massive clusters in the Milky Way, represents an ideal laboratory for studying the effects of high stellar density and ionizing radiation from massive stars on BD formation.}
   {This paper presents deep JWST/NIRCam observations of Wd2 in a large number of filters between 1.15 and 4.1 $\mu m$. Our analysis focuses on identifying and characterizing the BD population within the cluster.}
   {We carried out PSF photometry on deep JWST/NIRCam data obtained in 4 wide and 6 medium-band filters, using DOLPHOT. The resulting catalog was used to identify BD candidates in Wd2 through spectral energy distribution (SED) fitting with atmospheric models. The methodology was validated using synthetic JWST photometry of spectroscopically confirmed young BDs together with simulated populations of typical contaminants, allowing us to define an empirical selection criteria.}
   {The 50\% detection limit of our NIRCam catalog is $\sim$0.015-0.02 $M_\odot$, at the distance, age, and extinction of Wd2, providing the deepest view of a supermassive star cluster to date. The validation demonstrates that the adopted SED-fitting methodology reliably recovers young brown dwarfs while remaining robust against contamination from reddened evolved stars. Applying this procedure to Wd2, we identify 353 substellar candidates. Most candidates (301) lie above the 10 Myr isochrone in the Hertzsprung--Russell diagram consistent with cluster membership, which are defined as strong candidates. Comparison with a control field indicates a contamination level of $\lesssim$5\% for the strong candidate sample down to masses of $\sim$0.01--0.015~$M_\odot$. We identify 73 candidates exhibiting infrared excess indicative of circumstellar disks. These objects occupy the expected infrared-excess locus in de-reddened color diagrams and represent $27.7^{+3.4}_{-3.2}$\% of the candidates detected in F410M.
   }
   {These results demonstrate the capability of JWST/NIRCam medium-band photometry to identify robust samples of candidate brown dwarfs in distant, embedded, and crowded massive star clusters using imaging observations alone. The resulting catalog provides a well-characterized sample for future spectroscopic confirmation and subsequent studies of the substellar population of Wd2.}


   \maketitle
%

\section{Introduction}
\label{intro}

The launch of the James Webb Space Telescope (JWST) has opened many doors for the study of substellar objects - brown dwarfs (BDs) with masses too low to sustain stable hydrogen burning (M$<$0.075 $M_\odot$) and planetary-mass objects with no fusion at all (M$<$0.015 $M_\odot$). JWST's unprecedented sensitivity in the infrared gives us the chance to identify substellar populations  that were previously beyond reach. In particular, it has enabled the study of substellar objects in massive young clusters beyond 1 kpc with extreme environments \citep{guarcello25}, in low metallicity young clusters \citep{yasui24,andersen25}, in globular clusters \citep{gerasimov24},  and even in the Magellanic clouds \citep{zeidler24}. This capability sets the stage for comparative studies of substellar populations across a broad range of star-forming environments.

This is key to constraining formation scenarios at the bottom of the initial mass function \citep[IMF,][]{luhman12_review,muzic19}. While most researchers in the field agree that BDs predominantly form ‘like stars’, the detailed physics that sets the number of BDs is under debate. Most formation scenarios would imply that BDs numbers should be a function of environment. Moreover, at planetary masses, we also expect an additional population of ejected giant planets \citep{scholz22}. However, an unambiguous identification of this population remains elusive.

The JWST-enabled observations of BDs in distant clusters are not without new challenges. In particular, there is a clear need to find robust and innovative routines to identify BDs in crowded regions, in regions with high extinction, and in regions with many very bright stars. Efficient methods need to be established to distinguish between BDs in a cluster and other red sources in the background. Recent studies have explored several JWST-based approaches, including slitless spectroscopy \citep{langeveld24}, multi-band photometry \citep{defurio24}, and colour-magnitude diagrams using several bands, followed by selective spectroscopy \citep{luhman24_ic348}. The verification of the nature of candidate BDs in star forming regions is a topic for ongoing research - see for example the discussions on the substellar population in the Orion Nebula Cluster \citep[ONC, e.g.,][]{luhman24_onc_jwst,luhman25_onc_jwst_spec}. 

Once young BDs are identified, analyzing their additional properties is essential for testing competing formation scenarios and understanding how these processes vary across environments. Key diagnostics include disk presence and properties \citep[e.g.,][]{arabhavi25_sample1,damian25}, binarity \citep[e.g.,][]{kraus12,defurio22}, and the spatial and kinematic distribution of substellar populations \citep[e.g.,][]{joergens06,parker11,parker23}.

In this paper we use JWST/NIRCam to identify BDs in the young massive cluster Westerlund~2 (Wd2), using photometry in a wide range of filters across the 1.15-4.1 $\mu m$ range. Wd2 is a $\sim$1-2 Myr-old, massive young cluster located at a distance of $\sim$4.4 kpc in the RCW 49 star-forming complex \citep{zeidler15,maiz22}. It is among the most massive young clusters in the Milky Way, hosting numerous O-type stars and a total stellar cluster mass of $\sim$36,000 $M_\odot$ \citep{zeidler17}. With a stellar mass well above $10^4 M_\odot$, Wd2 belongs to the class of so-called supermassive star clusters \citep{portegies10}. Its massive stellar content, high stellar density, and strong feedback environment make it an ideal laboratory for testing how extreme conditions influence the properties of substellar objects located at the bottom of the IMF.

This paper is organized as follows: Section~\ref{observations} describes the NIRCam observations and data reduction; Section~\ref{photometry} details the PSF photometric analysis and the extraction of a catalog in all filters; Section~\ref{results_sed} presents the identification of substellar candidates via spectral energy distribution (SED) fitting; and Section~\ref{summary} summarizes the results and conclusions.

\section{Observations and data reduction}
\label{observations}

In this paper, we present JWST/NIRCam observations of Wd2, obtained as part of the Extended Westerlund 1 and 2 Open Clusters Survey (EWOCS) project\footnote{\url{https://Westerlund1survey.wordpress.com/}} (GO-3523, P.I.: Guarcello). EWOCS aims at studying how the harsh environments of supermassive star clusters shape the formation and early evolution of stars and planets. The observations consist of a single NIRCam pointing covering the core of Wd2 as well as a secondary cluster clump identified by \citet{zeidler15}. 

The observations were obtained on 2024-07-14, centered at $\alpha=10^\mathrm{h}23^\mathrm{m}49.994^\mathrm{s}$, $\delta=-57^\circ45'34.97''$, using the SHALLOW4 readout pattern, with 7 groups per integration, and one integration per exposure. We performed 4 subpixel dithers, resulting in a total integration time of 1460.201 s per filter. To characterize the contamination from foreground and background sources, we also observed a control field (CF) near Wd2 with identical filter coverage and observing strategy. These observations enable the statistical decontamination of the science field and the derivation of a reliable IMF of the cluster. The CF is centered at $\alpha=10^\mathrm{h}15^\mathrm{m}17.000^\mathrm{s}$, $\delta=-56^\circ59'42.00''$, and was observed on 2024-04-23. 

The observations include 16 filters spanning a set of wide, medium, and narrow bands. This broad filter-set enables a comprehensive study of both the stellar and substellar populations. In particular, several medium-band filters were selected to probe key spectral features in BD atmospheres, allowing their identification through purely photometric means, as will be the main focus of this paper (see Sect.~\ref{results_sed}). In this paper, we present results based on the wide and medium filters: F115W, F150W, F162M, F182M, F200W, F250M, F277W, F300M, F335M and F410M.

Data reduction was performed using the JWST pipeline version 1.16.0 and CRDS context 1298.pmap. In the Stage 1 of the pipeline (detector-level corrections), we set \texttt{suppress\_one\_group = False} to allow the use of a good zeroframe or first group for slope fitting in cases of saturation. Although this may reduce ramp-fit precision, it improves recovery of partially saturated stars and limits the impact of saturation-induced PSF streaks on the mosaics. The remainder of the data reduction was executed using default settings. This includes Stage 2 (calibration of individual exposures) and Stage 3 (production of final mosaics) of the JWST pipeline.

We opted not to apply a correction for the “streaking” artifacts caused by $1/f$ noise in the wide and medium filters. In this dataset, such noise is generally negligible compared to the dominant nebular background. We tested the JWST-adapted $1/f$ correction routine developed by C. Willott\footnote{\url{https://github.com/chriswillott/jwst}}, but found that it introduced image artifacts under all configurations considered.

\begin{figure*}[hbt!]
    \centering
    \includegraphics[width=\textwidth]{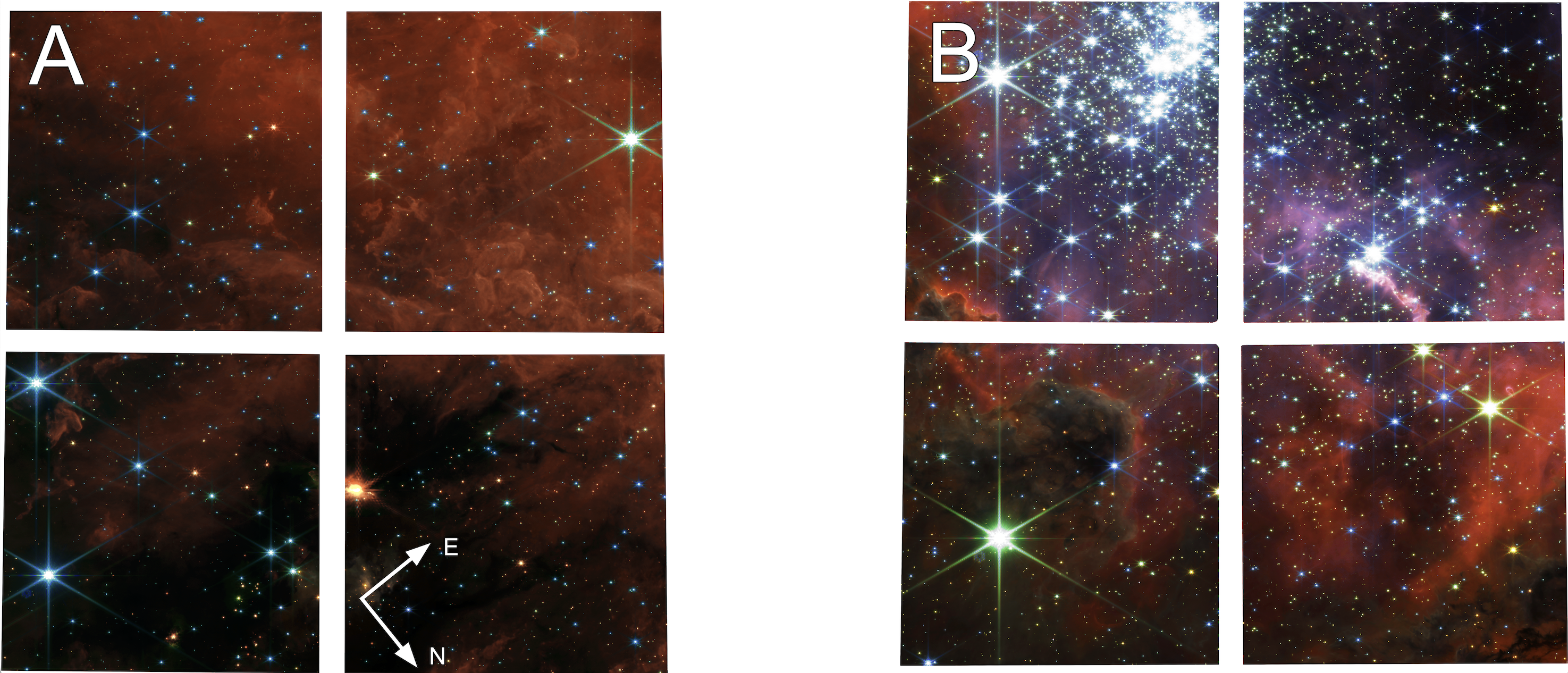}
    \caption{Color-composite of the NIRCam observations of Wd2  made with the F115W (blue), F150W (green) and F200W+F335M (red) filters. The field of view spans $5.1'\times2.2'$, corresponding to an area of $6.5\,\mathrm{pc} \times2.8\,\mathrm{pc}$ at a distance of 4.4 kpc. The image shows the eight NIRCam detectors, with Module A (left) and Module B (right) indicated by large letters in the upper left corner of each module. North and East are marked in the lower part of Module A.}
    \label{fig:rgb_wd2}
\end{figure*}

Fig.~\ref{fig:rgb_wd2} shows a color-composite image of Wd2. Module B (right panel) includes both primary cluster clumps. Detector B3 (top left quadrant) contains the central clump of Wd2, where a few saturated stars dominate the core, impacting completeness at faint magnitudes. Detector B1 (top right quadrant) contains the secondary clump, distinguished by a large, elongated nebulous structure—referred to as the ''sock'' \citep{zeidler18}—which is approximately 10$''$ ($\sim$0.21 pc at 4.4 kpc) in size. This feature lies within a roughly circular cavity of 17$''$ radius ($\sim$0.36 pc at 4.4 kpc), possibly excavated by stellar feedback \citep{zeidler18}. Detectors B2 and B4 (bottom quadrants) are dominated by diffuse emission and extended nebulae. Module A (left panel) covers a region adjacent to Wd2, not physically connected to the cluster, but rich in large-scale nebulosity. A detailed analysis of notable features across both modules—including globules, candidate photoevaporating structures, and outflows—based on the NIRCam narrow filters and MIRI observations will be presented in a forthcoming paper (Monsch et al. in prep).

\section{Photometry}
\label{photometry}

We performed PSF fitting photometry on the NIRCam observations of Wd2 using DOLPHOT \citep{Dolphin00, Dolphin16}. DOLPHOT performs photometry on the Stage 2 (\texttt{cal.fits}) images using the Stage~3 mosaics as a reference for source detection and image alignment. We used the DOLPHOT version released on February 4 2024, which includes the latest PSF libraries, Sirius-Vega zero points and \texttt{-etctime} pre-processing flag that adjusts the exposure times in the header to be in agreement with the exposure time calculator (ETC). The DOLPHOT PSF libraries were calculated using WebbPSF\footnote{\url{https://webbpsf.readthedocs.io/en/stable/index.html}} \citep{perrin12,perrin14} version 1.2.1. See \citet{weisz24} for a detailed description of how the DOLPHOT PSF library was built. We adopted the parameters recommended in \citet{weisz24} and the DOLPHOT NIRCAM manual\footnote{\url{http://americano.dolphinsim.com/dolphot/dolphotNIRCAM.pdf}}, which include prescriptions for PSF and sky fitting, aperture corrections, and image alignment strategies tailored for the short (SW) and long wavelength (LW) detectors.

Photometry was performed on all the sources above the pre-defined minimum signal-to-noise threshold (\texttt{sigPSF} parameter). This value is fixed at 5 as recommended in \citet{weisz24}. As expected, a significant fraction of the sources in the raw catalog are spurious, particularly detected along saturation spikes of bright stars and in nebulous regions. DOLPHOT also provides several photometry fit metrics that we used to define a set of selection criteria aimed at maximizing the completeness of real point sources while minimizing contamination. Starting from the cuts proposed for point sources in \citet{warfield23}, we optimized the thresholds in the F200W and F300M filters and subsequently applied them to the rest of SW and LW filters, respectively. We adopted the following cuts in the SW filters:

\begin{itemize}
    \item SNR=5-10, crowding$\leq$0.1, sharpness$^2 \leq$0.01
    \item SNR$\geq$10, crowding$\leq$0.3, sharpness$^2 \leq$0.01
    \item SNR$\geq$50, crowding$\leq$0.1, sharpness$^2 \leq$0.05
\end{itemize}

For LW filters, we used the same thesholds for the first two SNR ranges, while the third was relaxed to:
\begin{itemize}
    \item SNR$\geq$50, crowding$\leq$0.3, sharpness$^2 \leq$0.05
\end{itemize}

These cuts have been defined to perform a tighter selection at lower SNRs, where spurious detections dominate, and become looser at higher SNRs where real point sources can appear extended or blended due to nebulosity. Only $\sim$5\% of the entire catalog found by DOLPHOT meet these quality cuts, in agreement with previous studies in similarly crowded fields such as Westerlund~1 \citep{guarcello25} or the original JWST DOLPHOT paper \citep{weisz24}.

We obtained a single catalog containing the photometry in all wide and medium filters by cross-matching them with a maximum separation of 0.03$''$, corresponding to 1 and 0.5 pixel in the SW and LW channels, respectively. This cross-matching tolerance proved robust even when matching the bluest (F115W) and reddest (F410M) filters. In this work, we considered only the sources with detections in at least all F150W, F162M, F182M, and F200W filters (12554 sources). This selection ensures that each source in the catalog has measurements in these four key bands, but all available photometry for these sources is retained and used in the analysis. A detailed justification for the choice of these specific filters is provided in Sect.~\ref{results_sed}. The CF contains 10603 sources after applying the same detection constraint.

\subsection{Color magnitude diagram}

\begin{figure*}[hbt!]
    \centering
    \includegraphics[width=\textwidth]{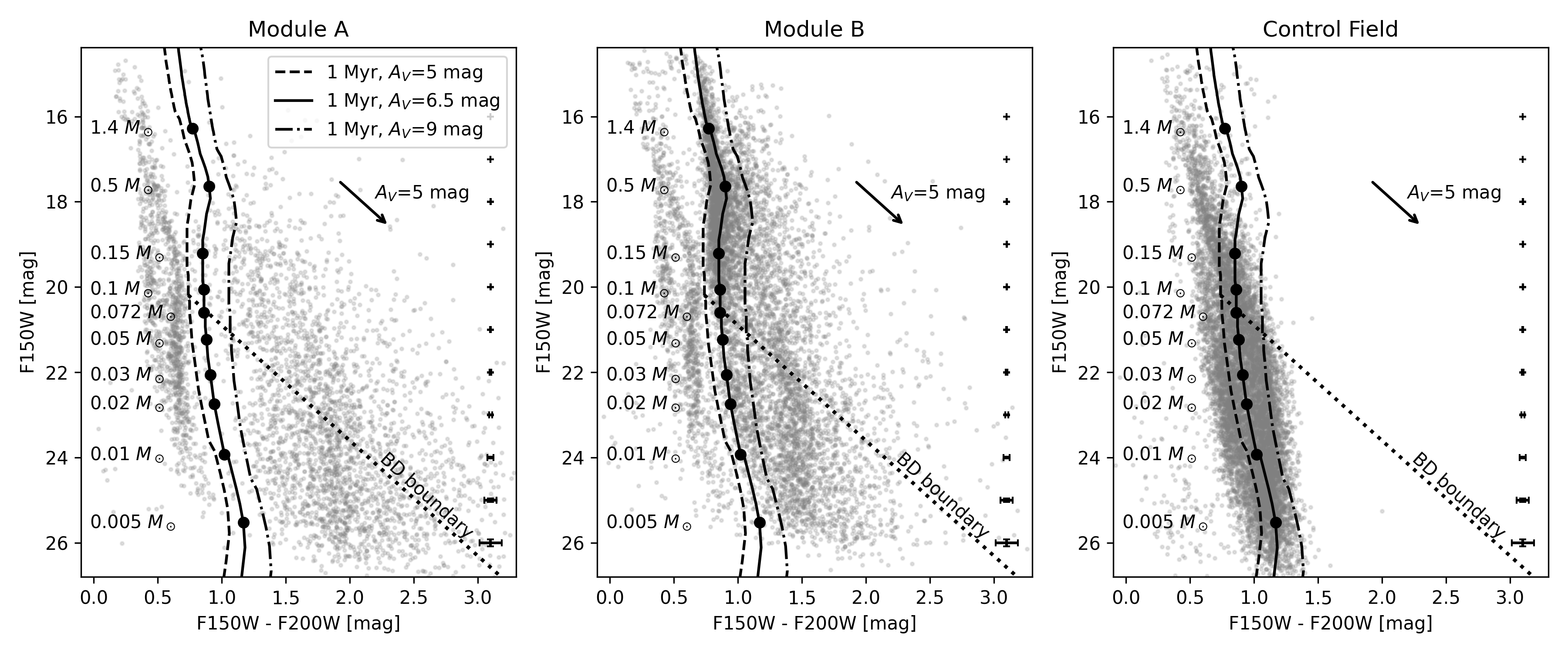}
    \caption{F150W$-$F200W vs. F150W CMD of the NIRCam observations of Wd2 of modules A (left panel), B (middle panel) and CF (right panel). The solid black line represents the stitched (see text for details) 1 Myr isochrone at a distance of 4.44 kpc and $A_V$=6.5 mag. The black dotted and dashed-dotted lines represent the same isochrone for $A_V$=5 and 9 mag, respectively. The black dotted line represents the star-BD boundary at different extinctions. Black errorbars at the right part of the each panel represent the mean errors at different magnitudes.}
    \label{fig:cmd_wd2}
\end{figure*}

Fig.~\ref{fig:cmd_wd2} presents the F150W$-$F200W vs. F150W color-magnitude diagram (CMD) of the Module A (left panel), Module B (central panel) and CF (right panel) observations. We overplot the 1 Myr \citet[hereafter BHAC15]{baraffe15} isochrone, shifted to a distance of 4.4 kpc \citep{maiz22} and an extinction of $A_V$ = 6.5 mag \citep{vargas13}. Age estimates for Wd2 in the literature lie in the range 1-2 Myr \citep{ascenso07,zeidler15}. Throughout this work we adopt 1 Myr as the fiducial age, as this isochrone provides a better representation of the observed cluster sequence at the high mass end of the CMD. We nevertheless explicitly test the impact of adopting a 2 Myr isochrone in all analyses that depend on evolutionary models, and report those results where relevant. Together with the adopted distance and extinction, this defines our reference set of cluster parameters. We adopt the extinction law from \citet{gordon23} with $R_V = 3.54$, as derived for Wd2 by \citet{wang24}. Below 0.01 $M_\odot$, where the BHAC15 models are no longer defined, we use the ATMO2020 isochrone \citep{phillips20} at the same age and extinction. The two sets of models align smoothly at the 0.01 $M_\odot$ transition, providing a consistent sequence across the substellar regime. At masses above 1.4 $M_\odot$, where BHAC15 models are also not available, we adopt PARSEC isochrones \citep{bressan12} at the same age and extinction. These models connect smoothly with the BHAC15 sequence in the CMD, allowing us to trace the cluster sequence consistently across the full stellar and substellar mass range. Throughout the rest of the paper, we use this stitched isochrone. Quantitative mass estimates are restricted to the BHAC15 mass range, while values outside this range are used only for approximate guidance (e.g., when discussing completeness limits).

The Wd2 stellar sequence is sharply defined in the Module B CMD, extending well into the BD regime (the substellar boundary, together with its extension along the reddening vector, is shown as a black dotted line) and is closely followed by the isochrone of the adopted age, distance and extinction. To illustrate the expected range of extinction among cluster members, we overplot additional isochrones corresponding to $A_V$ = 5 and 9 mag (dashed and dash-dotted lines, respectively). The majority of sources in Module B lie between these two isochrones, outlining the cluster sequence. In contrast, this sequence is virtually absent in Module A, which shows a strong paucity of sources along the same locus. The sharpness of the sequence in Module B is likely aided by the relatively low differential extinction across the cluster, owing to the cavity cleared by stellar feedback in the surrounding molecular cloud \citep{vargas13, zeidler15}. Both modules display two distinct foreground populations (F150W$-$F200W$\lesssim$0.75) and a more dispersed background population, with Module A sources appearing not only redder on average but also showing a wider scatter, consistent with stronger and more variable extinction. This color difference likely reflects the absence of a feedback-cleared cavity in Module A, resulting in higher average extinction toward background sources in that region. 

The CF exhibits a markedly different morphology from both modules. In the absence of extinction associated with the Wd2 cloud complex, the CMD sequence is confined to a relatively narrow locus in color. The foreground populations are less clearly separated, and the diagram is dominated by a smooth distribution of field stars. Based on these morphological differences, our analysis focuses on identifying substellar candidates in Module B, while using Module A and CF to estimate the level of contamination.

\subsection{Completeness}
\label{phot_completeness}

To quantify the detection limits of the observations, we performed artificial star tests using DOLPHOT. Artificial sources were injected one at a time to avoid self-crowding, sampling uniformly the magnitude range of interest and the full spatial extent. The position and magnitude of the artificial stars are then measured by DOLPHOT as if they were real sources. In total, we injected $\sim$10000 stars per filter and per field. A source was considered recovered if it was detected and satisfied the same photometric quality criteria adopted for the science catalog. Artificial star tests were performed independently in the four filters of interest (F150W, F162M, F182M, and F200W), and separately for Module~B, Module~A, and the CF. In addition, for Module~B we distinguish between the cluster core (inner 25$''$) and the outer region to assess the impact of crowding and saturation. We injected $\sim$3000 sources in the core area to ensure sufficient statistics.

\begin{figure}[hbt!]
    \centering
    \includegraphics[width=\textwidth/2]{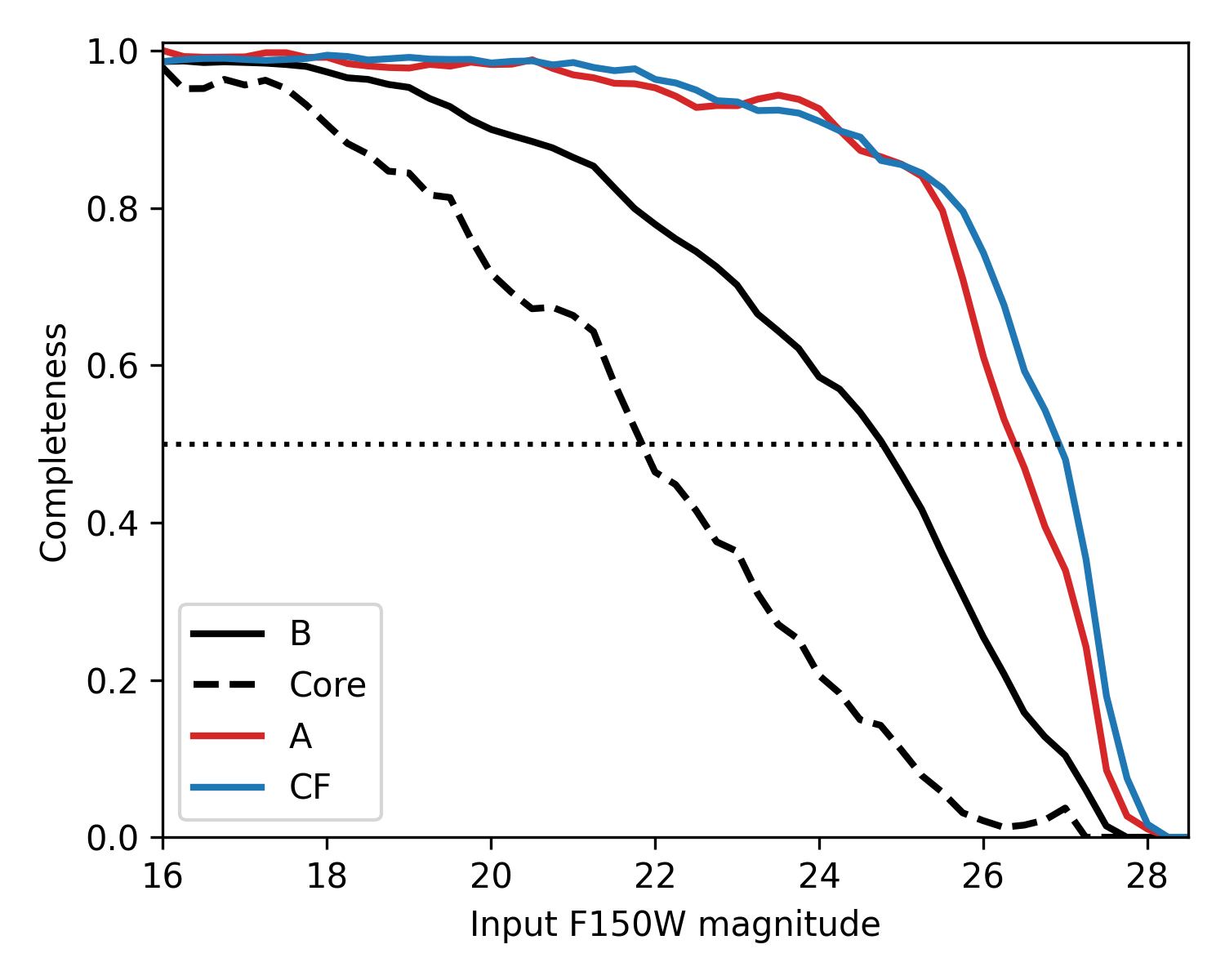}
    \caption{F150W completeness curves for the entire Module B (black solid line), Module A (red) and CF (blue). Black dashed line represents the completeness curve of the inner 25$''$ of the cluster in Module B.}
    \label{fig:comp_F150W}
\end{figure}

Fig.~\ref{fig:comp_F150W} presents the F150W completeness curve for the different fields. Module~B is significantly shallower than Module~A and the CF by more than one magnitude at the 50\% completeness level. Within Module~B, the cluster core (dashed line) shows a further decrease in the 50\% completeness relative to the outer region by two magnitudes, reflecting the effects of crowding and bright saturated stars.

The effective completeness of our catalog is governed by the shallowest filter among those required for detection. We find that F162M provides the most restrictive completeness limit, likely due to having the narrowest bandpass among the four filters inspected. The 50\% completeness magnitudes for the four filters and the Module B, Module A and the CF are listed in Table~\ref{tab:compl50}. For the adopted cluster parameters (1 Myr, $A_V = 6.5$ mag), the 50\% completeness level in the limiting filter corresponds to a mass of $\sim$0.015~$M_\odot$ in Module~B, and $\sim$0.021~$M_\odot$ if a 2 Myr isochrone is used.

\begin{table}
\caption{50\% completeness magnitude (top rows) and corresponding mass range for 1--2 Myr isochrones assuming $A_V=6.5$ mag (bottom rows) for the four key filters of interest.}
\label{tab:compl50}
\centering
\begin{tabular}{l c c c c}
\hline\hline
Filter & B & B$_{\rm core}$ & A & CF \\
\hline
F150W & 24.78 & 21.84 & 26.38 & 26.92 \\
 & 7--10 $M_\mathrm{Jup}$ & 35--41 $M_\mathrm{Jup}$ & 4--5 $M_\mathrm{Jup}$ & 3--4 $M_\mathrm{Jup}$ \\
\hline
F162M & 22.58 & 21.14 & 25.63 & 26.1 \\
 & 15--21 $M_\mathrm{Jup}$ & 40--45 $M_\mathrm{Jup}$ & 3--4 $M_\mathrm{Jup}$ & 3--4 $M_\mathrm{Jup}$ \\
\hline
F182M & 22.96 & 20.83 & 24.76 & 26.0 \\
 & 12--16 $M_\mathrm{Jup}$ & 44--49 $M_\mathrm{Jup}$ & 5--7 $M_\mathrm{Jup}$ & 3--5 $M_\mathrm{Jup}$ \\
\hline
F200W & 22.48 & 20.7 & 24.7 & 25.9 \\
 & 13--18 $M_\mathrm{Jup}$ & 41--46 $M_\mathrm{Jup}$ & 4--6 $M_\mathrm{Jup}$ & 3--4 $M_\mathrm{Jup}$ \\
\hline
\end{tabular}
\end{table}

\section{Identification of brown dwarfs in Westerlund 2 through SED fitting}
\label{results_sed}

The observations of Wd2 have been performed in a rich suite of filters that include medium filters designed to probe spectral features characteristic of very low-mass stars and BDs. Notably, the F162M and F182M filters cover the prominent H-band peak around $\sim$1.67 $\mu$m \citep{scholz12}, surrounded by water absorption at both sides at $\sim$1.4 $\mu m$ and $\sim$1.8 $\mu m$ respectively. The prominence of this feature increases with decreasing mass or $T_\mathrm{eff}$ \citep{cushing05} and becomes distinctly triangular at younger ages \citep{lucas06,allers13,almendros22}. Our filter-set also includes additional medium filters (F250M, F335M, F410M) sensitive to other molecular features such as $H_2O$ and $CH_4$ absorption, which are similarly enhanced in very low-mass BDs. 

\begin{figure}[hbt!]
    \centering
    \includegraphics[width=\textwidth/2]{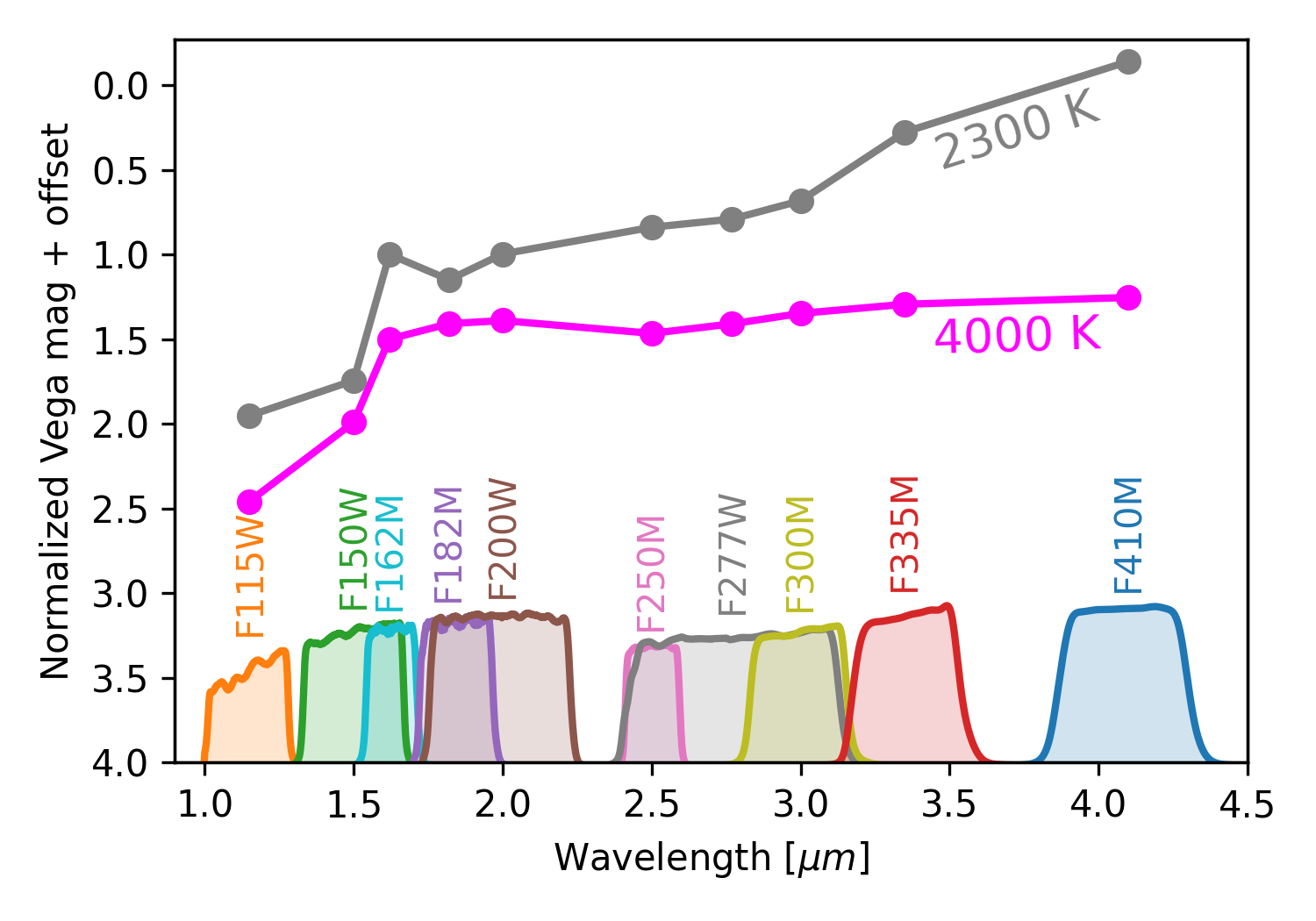}
    \caption{Transmission curve of all wide and medium filters available in the Wd2 NIRCam observations presented here. We also plot two SEDs with $\log g=4$ and $T\mathrm{_{eff}}$ of 2300 and 4000 K from the BT Settl models.}
    \label{fig:sed_bt}
\end{figure}

We perform synthetic photometry of the BT-Settl models \citep{allard11} covering a broad range of $T_\mathrm{eff}$ values (1100 to 8000 K), accross all the wide and medium filters we have available. Fig.~\ref{fig:sed_bt} presents the synthetic SED of two BT-Settl models with $T_\mathrm{eff} =$2300 K and 4000 K and $\log g=4$, a value appropriate for young stars and BDs. These SEDs illustrate the characteristic shapes of a substellar object and a low-mass star, respectively, along with the transmission curves of the medium and wide filters. A key feature observed in the SED of the $T_\mathrm{eff} = 2300$ K model is a pronounced dip between the F162M, F182M, and F200W bands. This dip becomes noticeable around $T_\mathrm{eff} \sim 3000$ K and is attributed to the onset of strong $H_2O$ absorption bands, as described above. The combination of this diagnostic with the redder photometric colors at longer wavelengths provides a powerful means to constrain the substellar nature of an object. This is the main reason why the analysis in this paper focuses on the subset of sources with detections in the F150W, F162M, F182M, and F200W filters, as introduced in Sect.~\ref{photometry}.

\subsection{SED fitting setup}
\label{results_sed_setup}

We fit the observed NIRCam SEDs using the BT-Settl atmospheric models, as described above, between 1100 and 8000 K with a step of 100 K. Extinction, $A_V$, is included as a free parameter and allowed to vary between 0 and 30 mag in steps of 0.25 mag. After applying extinction, the synthetic model photometry is normalized to the median observed SED of each target. We find that the fits are not strongly sensitive to variations in surface gravity; thus, we fix $\log g = 4$. The best-fit model is identified by minimizing the reduced $\chi^2$ statistics between the observed SED and the synthetic photometry over the full grid of $T_\mathrm{eff}$ and $A_V$, defined as:

\begin{equation}
    \chi^2=\frac{1}{N-m} \sum_{i=1}^{N} (O_i-T_i)^2
\end{equation}

where $O$ is the object SED, $T$ the model SED, $N$ the number of photometry measurements available, and $m$ the number of fitted parameters ($m$=2). We perform two separate SED fits for each target. The first fit uses the full wavelength coverage in order to exploit the overall spectral shape and wide-wavelength leverage on extinction. The second fit is restricted to the 1.15-2 $\mu$m range, isolating the water-band diagnostics that are most sensitive to temperature in substellar objects and largely insensitive to disk emission. At ages between 1 and 5 Myr, the stellar-BD boundary lies at approximately 2900 K, according to the BHAC15 evolutionary isochrones. We consider as substellar candidates those sources with best-fit $T_\mathrm{eff} \leq 3000$ K in both SED fits, and adopt the $T_\mathrm{eff}$ and $A_V$ values derived from the full SED fit, unless infrared excess affects the long-wavelength photometry, as discussed below.

\subsection{Validation of  substellar candidate SED-based selection}
\label{results_sed_validation}

In this Section we validate the SED fitting setup using a set of known young brown dwarfs and several classes of potential contaminants. Given the relative novelty of JWST, there are limited photometric observations of known young BDs in the specific filter-set required to validate our methodology. \citet{wang24} used 627 JWST/NIRSpec PRISM spectra from four different observing programs to derive the extinction law toward Westerlund 2. Of these, only program 2640 includes sources within the field of view of our JWST/NIRCam observations. However, when placed in the F150W$-$F200W vs. F150W CMD, none of these sources lies below the hydrogen-burning limit as defined by the stitched isochrone. Therefore, this sample does not include substellar objects and cannot be used to validate our methodology.

As an alternative, we use 1-5 $\mu$m JWST/NIRSpec PRISM spectra of bona fide young BDs in the ONC from \citet{luhman24_onc_spec} and late-type ($\geq$M9) young BDs from \citet{damian25}. These datasets include young ($<$5 Myr) objects with spectral types (SpTs) ranging from M6.5 to L4, which corresponds to $T_\mathrm{eff}$ between 1600 K and 2900 K, based on established SpT-$T_\mathrm{eff}$ relations \citep{luhman03_spt_teff,sanghi23}. As such, it spans the full $T_\mathrm{eff}$ range expected for the substellar population in Wd2. To ensure reliable fits, we exclude sources exhibiting strong HI emission lines, as these features are significantly broadened due to the low spectral resolution of the PRISM data. We obtain synthetic photometry from the spectra and apply the SED fitting procedure described in Sect.~\ref{results_sed_setup}.

\begin{figure*}[hbt!]
    \centering
    \includegraphics[width=\textwidth]{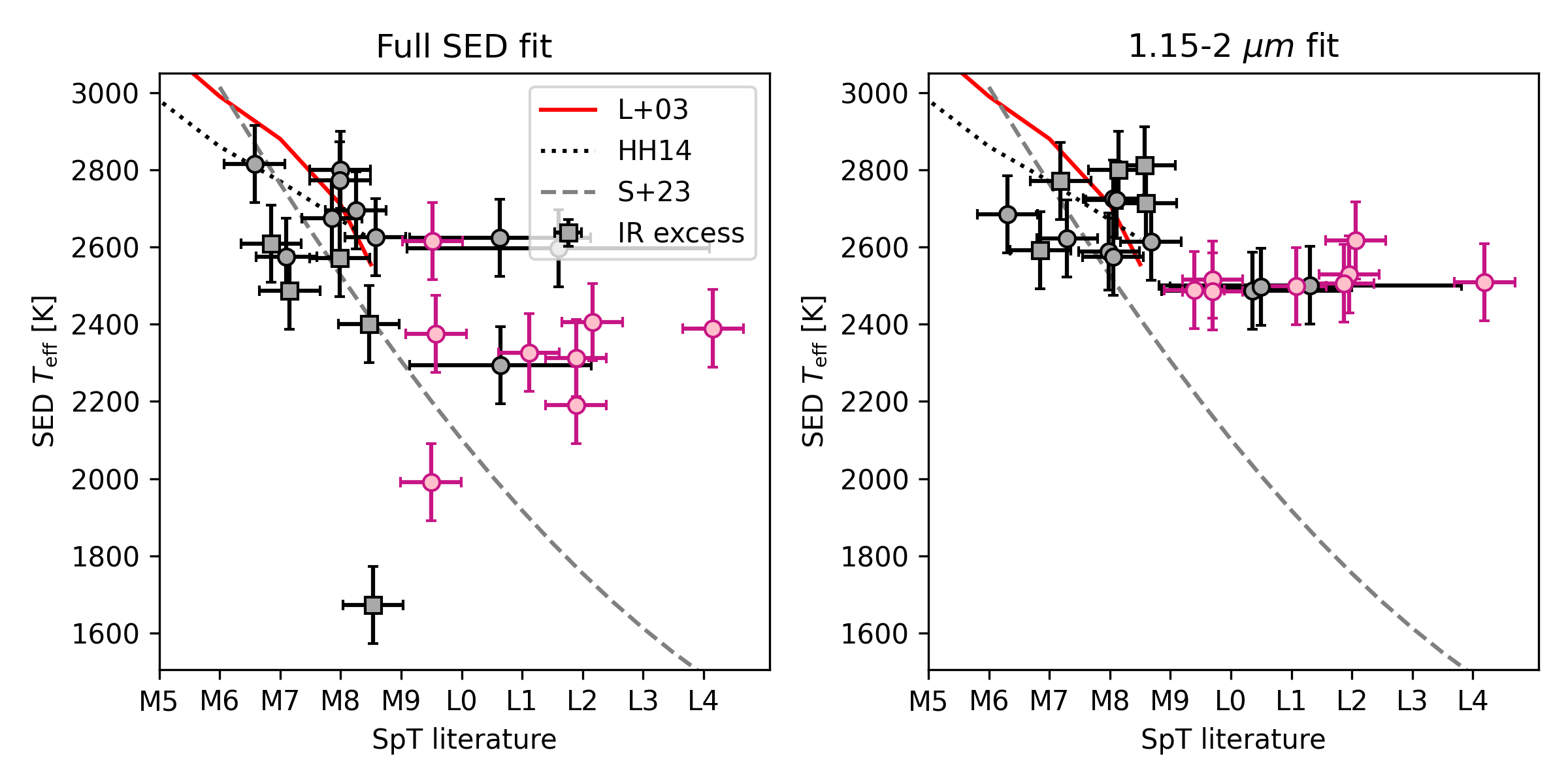}
    \caption{Comparison between the SpT values from \citet[black,][]{luhman24_onc_spec} and \citet[pink,][]{damian25}, and the $T_\mathrm{eff}$ values derived from SED fitting for the young late-type sources in our validation set. The left panel shows $T_\mathrm{eff}$ from the full SED fit, while the right panel shows results from the fit limited to the 1.15-2~$\mu$m range. Squares denote sources showing strong infrared excess \citep{luhman24_onc_spec}. The uncertainty in $T_\mathrm{eff}$ corresponds to the model grid step (100 K). For reference, we include the SpT-$T_\mathrm{eff}$ relations from \citet{luhman03_spt_teff} (red solid line), \citet{herczeg14} (black dotted line), and \citet{sanghi23} (gray dashed line). A small random offset ($\leq$0.2 spectral subtypes and $\leq$30~K in $T_\mathrm{eff}$) has been applied to individual points for clarity.}
    \label{fig:sed_comp_l24}
\end{figure*}

All sources in the sample meet the substellar candidate requirements and have $T_\mathrm{eff} \leq 3000$ K in both the full and restricted SED fits. When comparing the derived temperatures to the SpTs reported in \citet{luhman24_onc_spec} and \citet{damian25}, we find good agreement with established SpT-$T_\mathrm{eff}$ relations for SpTs $\leq$ M9 (black and pink circles in Fig.~\ref{fig:sed_comp_l24}, respectively). At SpTs $>$ M9, where SpT is known to be highly degenerate with extinction \citep{luhman17_tau}, we derive somewhat higher temperatures than expected from the reported SpTs, together with systematically higher $A_V$ values. This effect is more pronounced in the restricted fit (right panel), where most $>$M9 sources converge toward best-fit temperatures around 2500 K. It is also important to note that SpT-$T_\mathrm{eff}$ conversions in this regime remain uncertain, since current calibrations are primarily based on nearby young moving groups \citep[10--100 Myr;][]{filippazzo15,sanghi23}. Overall, the agreement between the fitted temperatures and the expected SpT-$T_\mathrm{eff}$ relations supports the reliability of the adopted fitting methodology over the full temperature range expected for the Wd2 substellar population.

\begin{figure}[hbt!]
    \centering
    \includegraphics[width=\textwidth*10/21]{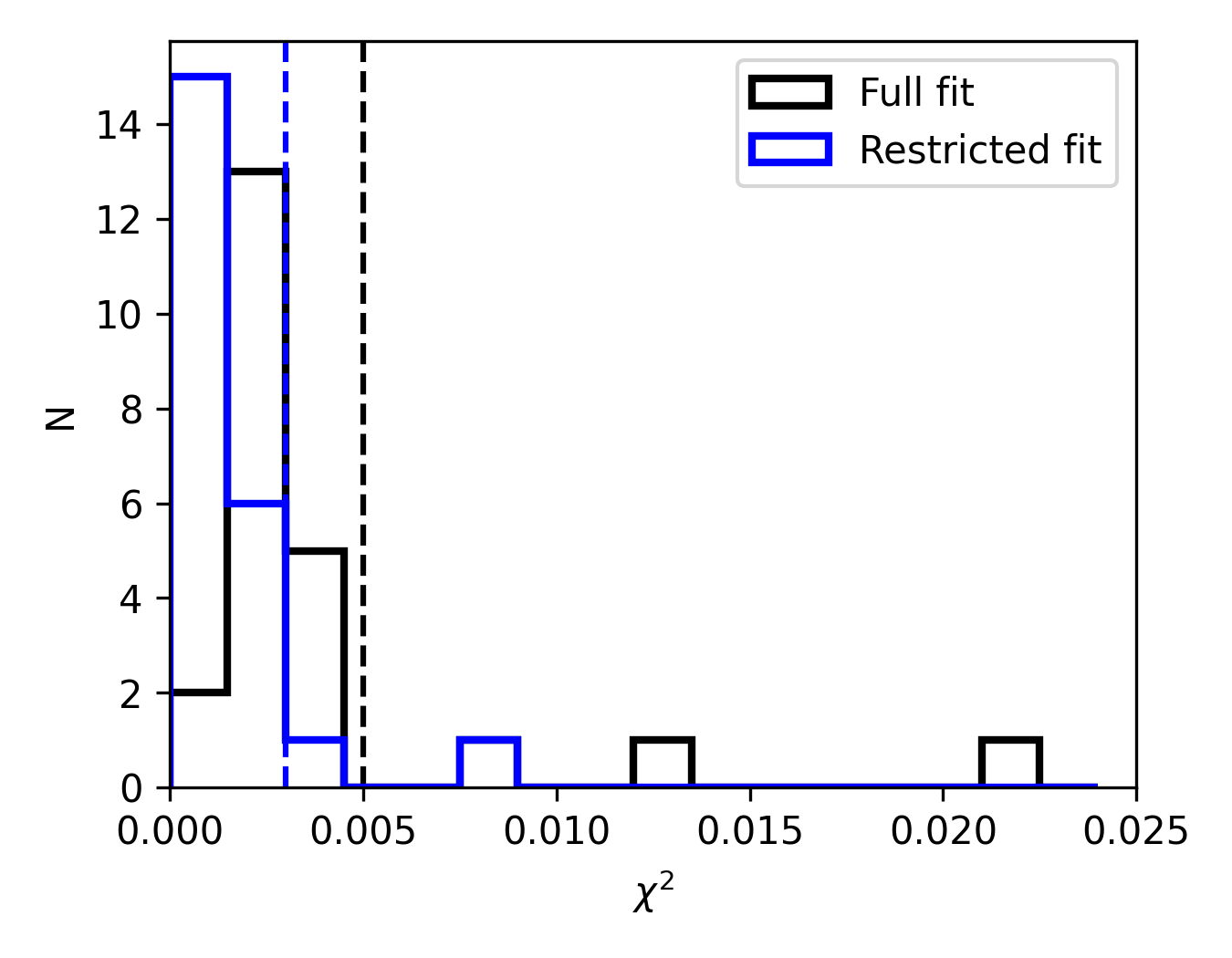}
    \caption{$\chi^2$ distribution of the best-fit of the bona fide young brown dwarf validation sample in the full SED (black) and restricted SED fits (blue). The adopted $\chi^2$ thresholds are shown with a black and blue vertical dashed lines for the full and restricted SED fits, respectively.}
    \label{fig:chi2_dist_validation}
\end{figure}

The validation sample also allows us to empirically define the $\chi^2$ thresholds to be adopted in the identification of substellar candidates in Wd2. Fig.~\ref{fig:chi2_dist_validation} shows the distribution of best-fit $\chi^2$ values for the bona fide young BD sample in both the full (black) and restricted (blue) SED fits. The vast majority of validation targets are recovered with $\chi^2 \leq 0.005$ for the full fit and $\chi^2 \leq 0.003$ for the restricted fit, and we therefore adopt these values as the criteria for reliable SED fits. The three sources that do not satisfy the full SED criterion correspond to objects with clear infrared excess emission, for which the longer-wavelength emission cannot be reproduced by purely photospheric models. Similarly, the two sources that do not satisfy the restricted-fit criterion show evidence of excess emission already at wavelengths around 2~$\mu$m, affecting even the restricted fit. This confirms that disk-bearing substellar objects can fail the full SED $\chi^2$ criterion despite being genuine young BDs. Therefore, when applying the selection to Wd2, we additionally inspect sources satisfying the temperature requirements and the restricted-fit $\chi^2$ criterion but failing the full SED criterion, allowing us to recover candidates where infrared excess affects the 
long-wavelength photometry.

In addition to the best-fit solutions, we explored the full $\chi^2$ surface over the $T_\mathrm{eff}$--$A_V$ grid for each source. Representative examples of these $\chi^2$ maps for several validation objects are shown in Fig.~\ref{fig:chi2_map}. The maps illustrate the expected degeneracy between $T_\mathrm{eff}$ and $A_V$, where different combinations of temperature and extinction can reproduce similar SED shapes. For most validation targets the minimum $\chi^2$ region remains localized around the best-fit solution.

We use these $\chi^2$ maps to evaluate the robustness of the stellar/substellar classification. For each source, we define all $T_\mathrm{eff}$--$A_V$ pairs satisfying the adopted $\chi^2$ thresholds as acceptable solutions, and quantify the fraction of those solutions located above or below the adopted substellar boundary at $T_\mathrm{eff}=3000$ K. These ranges provide an empirical measure of the extent of the temperature--extinction degeneracy within the adopted fitting criteria. In the full SED fits, none of the bona fide young BD validation targets have more than 50\% of acceptable solutions above the substellar boundary. In the restricted fits, only one source (\#126), which also presents an excess-dominated SED, has more than 50\% of acceptable solutions above 3000 K. This indicates that the adopted substellar classification is robust against the $T_\mathrm{eff}$--$A_V$ degeneracy.

A subset of the sources in \citet{luhman24_onc_spec} show strong infrared excesses attributed to protoplanetary disks. These objects illustrate why the restricted fit is required in addition to the full SED fit. Disk emission can bias the longer-wavelength photometry and drive the full SED fit toward artificially low temperatures. By restricting the fit to the 1.15--2.0~$\mu$m range, where the emission remains predominantly photospheric, we recover temperatures that are more consistent with the expected spectral types while still retaining the ability to identify the sources as substellar candidates. The most notable example is source \#85 from \citet{luhman24_onc_spec}, whose full SED fit yields a very low temperature due to strong excess emission, whereas the restricted fit recovers a significantly higher temperature of $\sim$2800 K, more consistent with other sources of comparable SpT. 

We assess whether post-main sequence stars could mimic the SEDs of substellar objects. To this end, we generate synthetic photometry for post-main sequence stars of various masses, ages, and evolutionary stages using the PARSEC evolutionary models \citep{bressan12}. In particular, we select models in the subgiant, red giant, and core He-burning phases, with ages between 1 and 10 Gyr and $T_\mathrm{eff}$ between 3100 and 8000 K. In observations of a young cluster, such stars are expected to appear as strongly reddened background sources, so we artificially redden their SEDs by a wide range of extinction. All post-main sequence models are best fit by temperatures above 3500 K and would therefore not be classified as substellar candidates, irrespective of the extinction applied. Furthermore, none of these models has more than 50\% of acceptable solutions below the adopted substellar boundary in either the full or restricted fits. This demonstrates that extinction alone does not drive evolved stars into the substellar regime through the $T_\mathrm{eff}$--$A_V$ degeneracy.

\begin{figure}[hbt!]
\centering
\includegraphics[width=\textwidth*10/21]{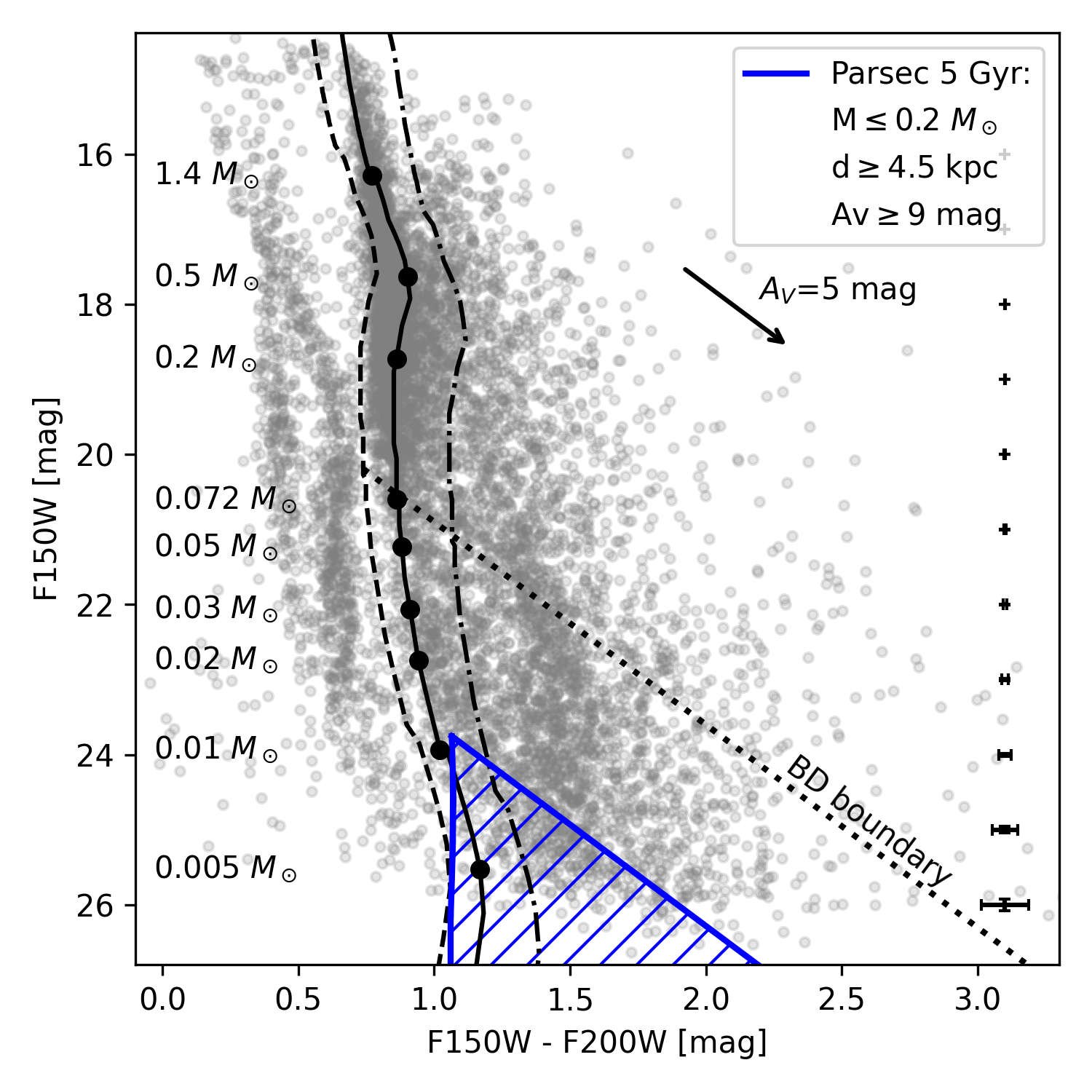}
\caption{F150W$-$F200W vs. F150W CMD following the same formatting of Fig.~\ref{fig:cmd_wd2}. The blue hatched area represents the area where background reddened main sequence dwarfs are expected to be located (see text for details).}
\label{fig:app_cmd_reddened}
\end{figure}

We also evaluate whether reddened main sequence stars could be classified as Wd2 substellar candidate members. We use PARSEC evolutionary models at an age of 5 Gyr to simulate field dwarfs and fit stars with masses between 0.09 and 1 $M_\odot$ across a wide range of extinction values. As expected, field dwarfs with intrinsic temperatures $\leq3000$ K can be classified as substellar by the SED-fitting procedure, since their photospheric temperatures are genuinely within the adopted substellar regime. In the full SED fits, all models with intrinsic $T_\mathrm{eff}\leq3000$ K have more than 50\% of acceptable solutions below the adopted boundary. In contrast, in the restricted fit only two models ($T_\mathrm{eff}=$ 2774 and 2988 K) satisfy this condition. This reflects the fact that the restricted fit is mildly sensitive to the gravity-dependent shape of the H-band peak. As a result, some intrinsically cool field dwarfs are rejected by the restricted fit despite having temperatures below 3000 K. Importantly, no hotter main-sequence model is preferentially classified as substellar, irrespective of the extinction applied. Thus, background reddened dwarfs could, in principle, be misidentified as BD candidates. However, since such sources would lie behind Wd2 (i.e., at $d \gtrsim 4.5$ kpc and $A_\mathrm{V} \gtrsim 9$ mag), they occupy a narrow region of CMD space (see blue hatched area in Fig.~\ref{fig:app_cmd_reddened}). This strongly limits their potential contribution to contamination.

Overall, these tests show that the adopted SED-fitting methodology reliably identifies young substellar objects as sources with $T_\mathrm{eff}\leq3000$ K, while remaining robust against contamination from reddened evolved stars. The only stellar contaminants that can naturally satisfy the temperature criterion are intrinsically cool field dwarfs, whose contribution is further reduced in the restricted fit and constrained by their limited location in CMD space. Moreover, artificially reddening both the field-dwarf and post-main-sequence models over a wide range of extinction values does not increase the number of contaminants classified as substellar, demonstrating that extinction itself does not bias the selection toward lower temperatures. The analysis of the acceptable $\chi^2$ solutions further shows that the $T_\mathrm{eff}$--$A_V$ degeneracy only introduces ambiguity for a small number of infrared excess-dominated sources, while the classification of the vast majority of bona fide young brown dwarfs remains robust.

\subsection{Identification of substellar candidates in Westerlund 2}
\label{results_sed_wd2}

We now apply the validated SED-fitting procedure to the NIRCam photometry of Wd2 in Module B. In addition to selecting the sources with the best-fit $T_\mathrm{eff} \leq 3000$ K, we require them to be located below the theoretical BD boundary in the F150W-F200W vs. F150W CMD (black dotted line in Fig.~\ref{fig:cmd_wd2}), ensuring consistency with substellar luminosities at the distance and age of the cluster. This luminosity cut also removes the brightest substellar binaries with high mass ratios. Assuming a binary fraction similar to that of the ONC ($\sim$12\%; \citealt{defurio22}), and considering that only systems near the substellar limit ($\sim$0.05-0.075$M_\odot$) would be excluded by this criterion, we estimate that this effect would cause us to miss less than 10\% of the true substellar population. The separation between foreground and cluster sources is clear in the F150W-F200W vs. F150W CMD, allowing us to reliably avoid foreground contamination. We further exclude sources more than 3$\sigma$ bluer than the $A_V = 5$ mag 1 Myr isochrone in this CMD (black dashed line in Fig.~\ref{fig:cmd_wd2_cands}). Of the 8324 sources in the Module B catalog, 2599 satisfy these CMD-based selection criteria and are subsequently fitted with the atmospheric model grid. Among them, 590 have best-fit temperatures of $T_\mathrm{eff}\leq3000$ K in both the full and restricted SED fits. Applying the SED fit empirical $\chi^2$ thresholds derived from the validation sample, yields 308 candidate substellar members.

\begin{figure}[hbt!]
    \centering
    \includegraphics[width=\textwidth/21*10]{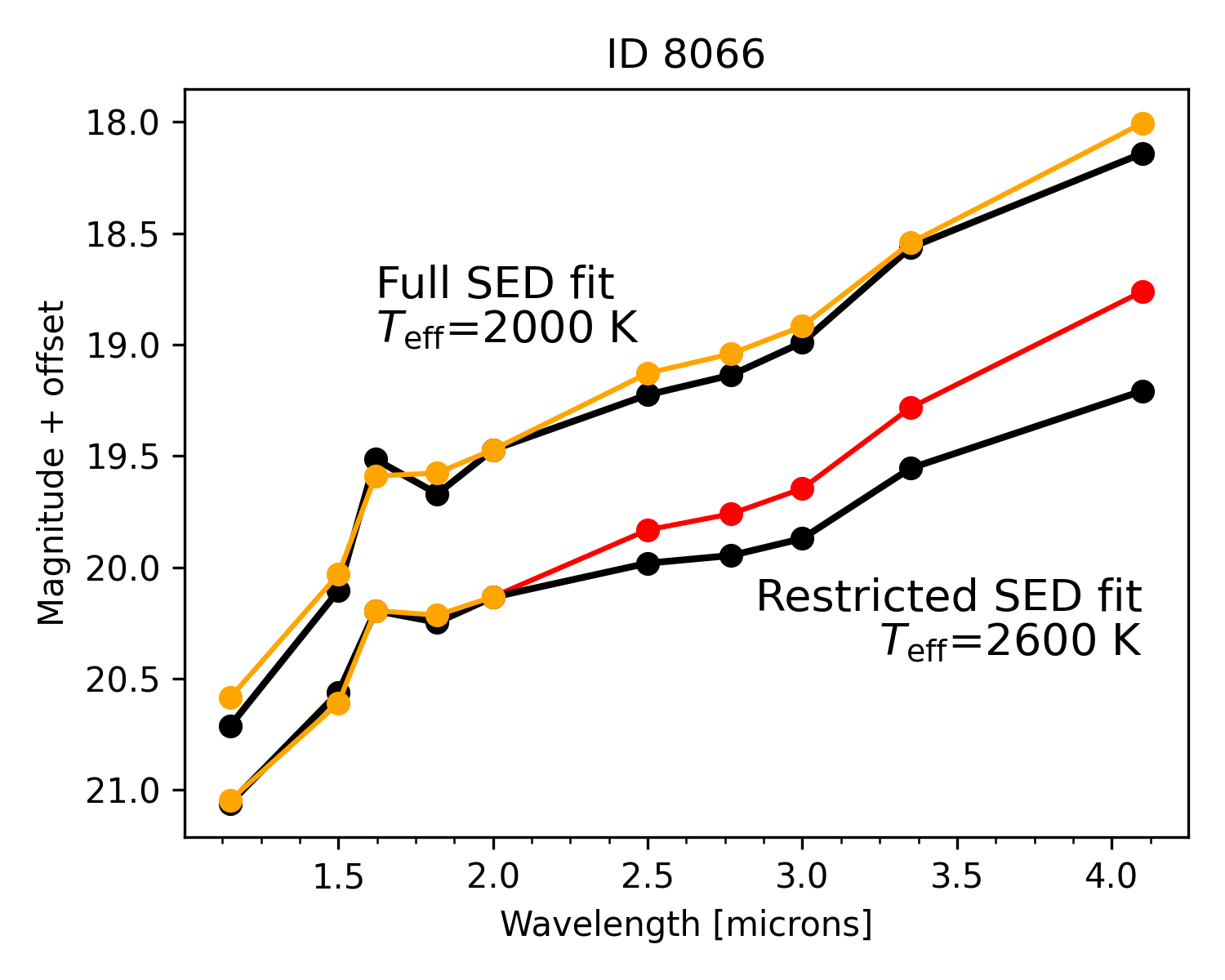}
    \caption{De-reddened SED of disk-bearing substellar candidate 8066 (orange line) together with the best fit model (black) from the full SED fit (top), and restricted SED fit (bottom). Red circles in the restricted target SED fit represent the model magnitudes not included in the fitting that are affected by infrared excess.}
    \label{fig:sed_cand_disk}
\end{figure}

\begin{figure}[hbt!]
    \centering
    \includegraphics[width=\textwidth/21*10]{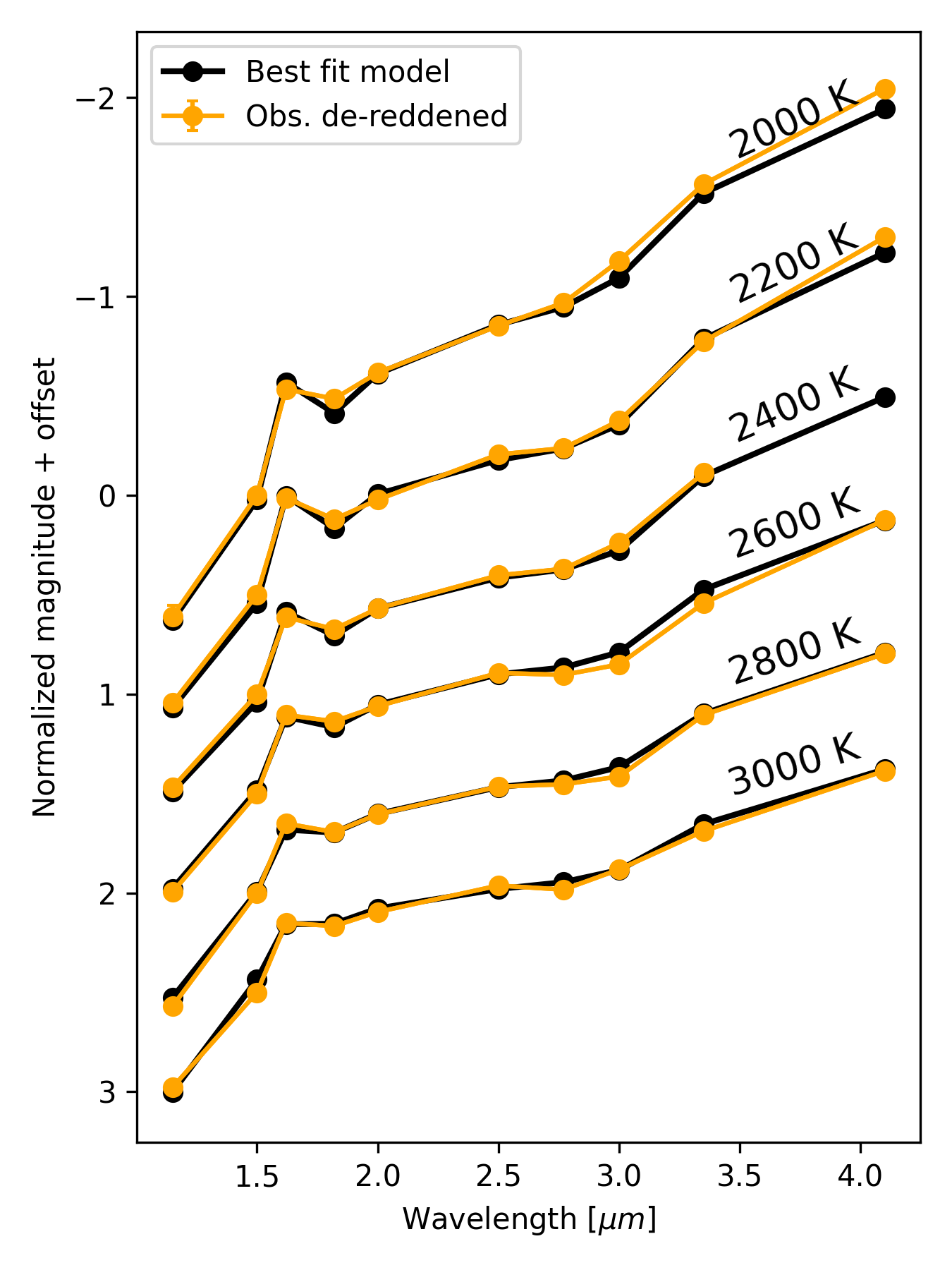}
    \caption{SED of six of the identified candidates in Sect.~\ref{results_sed_wd2} with increasing $T_\mathrm{eff}$ (orange line) together with their best fit model (black line). The observed SEDs have been de-reddened by the best fit $A_V$, as well as normalized and offset from each other.}
    \label{fig:sed_wd2}
\end{figure}

As demonstrated by the validation sample, disk emission primarily affects the longest wavelengths, and genuine disk-bearing substellar objects may fail the $\chi^2$ criterion in the full SED fit while remaining well reproduced in the restricted fit that probes the water absorption features. To recover such sources, we visually inspect sources with $T_\mathrm{eff} \leq 3000$ K in both fits that satisfy the restricted $\chi^2$ threshold but not the full-SED one. Objects with infrared excess are straightforward to recognize, as they display clear excess emission at $\lambda \gtrsim 2~\mu$m relative to the best-fit photospheric model from the restricted fit (see Fig.~\ref{fig:sed_cand_disk}). We identify 45 objects with infrared excess through this procedure and include them in the candidate list, yielding a final sample of 353 Wd2 substellar candidates. Within this final sample, we then assess the presence of infrared excess for all objects. In addition to the sources recovered through the procedure above, we identify 28 candidates that satisfy the $\chi^2$ requirements in both fits but nevertheless show excess emission at long wavelengths. Altogether, this results in 73 Wd2 substellar candidates with evidence of disks. For all these sources that present infrared excess, we adopt the $T_\mathrm{eff}$ and $A_\mathrm{V}$ values derived from the restricted fit, for the remaining we use the full SED fit. We adopt the range of acceptable solutions in the adopted fit (i.e., solutions meeting the $\chi^2$ criteria) as the uncertainty range in $T_\mathrm{eff}$ and $A_V$. This range is intended to capture the effect of the $T_\mathrm{eff}$--$A_V$ degeneracy over the discrete model grid and is therefore conservative in cases where the acceptable $\chi^2$ region includes extended low-significance tails. In Fig.~\ref{fig:sed_wd2}, we present example SEDs for six Wd2 candidates spanning a range of $T_\mathrm{eff}$ from 2000 to 3000 K (orange), de-reddened using the best-fit extinction, alongside their best-fit BT-Settl models (black).

We evaluated the robustness of the derived temperatures against two potential sources of systematics: extinction law and photometric errors. First, varying the extinction law within the range $R_V$=3-4 results in changes in $T_\mathrm{eff} \leq$100 K (the model grid step) for more than 95\% of the sample. Second, Monte Carlo realizations of the photometry using the errors yield changes in $T_\mathrm{eff} \leq$100 K for 98\% of the sources brighter than F150W = 23 mag, and for about 88\% of the fainter objects. Finally, inspection of the acceptable $T_\mathrm{eff}$--$A_V$ solutions shows that only a small number of sources have ambiguous stellar/substellar classifications, and that such cases are concentrated near the adopted $T_\mathrm{eff}=3000$ K boundary or among sources with limited wavelength coverage. A comparable number of borderline sources not selected as substellar also admit acceptable substellar solutions. Therefore, the $T_\mathrm{eff}$--$A_V$ degeneracy does not introduce a significant one-sided bias in the final candidate sample. Since our selection criterion depends only on whether $T_\mathrm{eff}$ is above or below 3000 K, these uncertainties have a negligible impact on the candidate classification.

To further assess whether the SED-fitting behavior of the Wd2 candidates is consistent with that of the validation sample, we compared the topology of the $\chi^2$ maps of both samples. The $\chi^2$ maps of the candidate sample exhibit the same qualitative topology as those of the validation sample (see Figs.~\ref{fig:chi2_map} and \ref{fig:chi2_map_cands}), consisting of a well-defined minimum together with an elongated $T_\mathrm{eff}$--$A_V$ degeneracy region. This demonstrates that the SED-fitting behavior is consistent between the validation and Wd2 samples.

\subsection{HR diagram and contamination estimation}
\label{results_sed_cont}

We perform the same analysis on the Module A and CF observations, where we expect little-to-no presence of Wd2 members. Out of the 4230 and 10603 sources in the Module A and CF catalogs, respectively, with valid photometry in the F150W, F162M, F182M and F200W filters, we identify 84 and 385 substellar candidates.  We verified that the lower extinction of the CF does not limit the identification of substellar candidates by artificially reddening the CF sources over a range of additional extinction values comparable to those observed toward Wd2 and reapplying the full selection procedure. The number of selected substellar candidates remains unchanged, indicating that extinction does not artificially increase the number of contaminants identified by the SED fitting. This is consistent with the validation tests presented in Appendix~\ref{results_sed_validation}, where reddening field and post-main-sequence models over a wide range of extinction values did not increase the number of objects classified as substellar candidates. 

\begin{figure*}[hbt!]
    \centering
    \includegraphics[width=\textwidth]{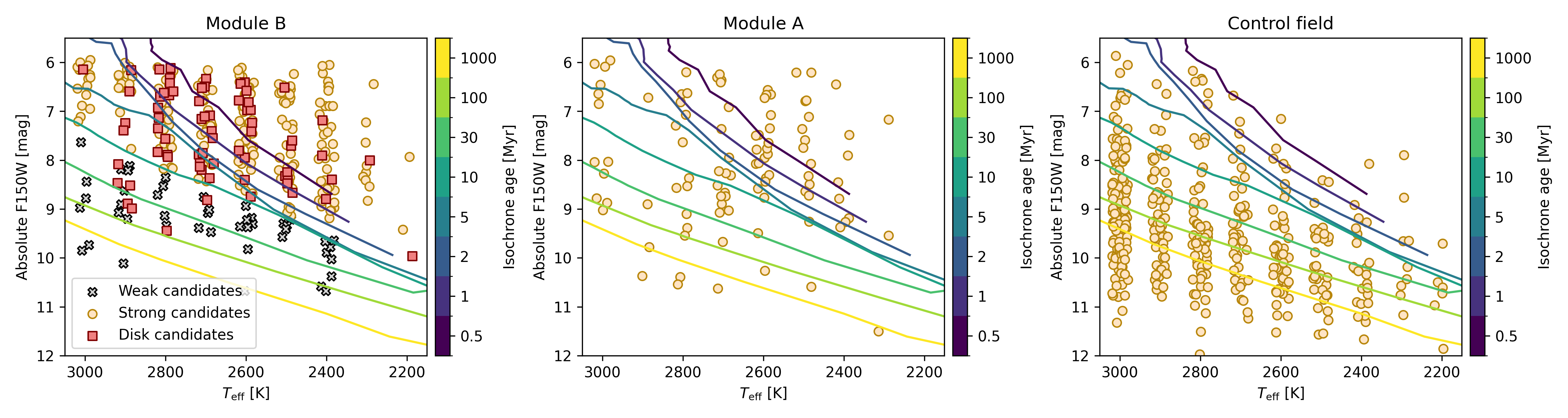}
    \caption{HRD of SED fitting substellar candidates identified in Module B (left panel), Module A (middle panel) and CF (right panel). Sources located below the 10 Myr isochrone (i.e., weak candidates) are shown with black crosses, those with infrared excess with red squares and the rest with orange circles. A random $\pm$20 K oﬀset was applied to all sources for clarity. Colored lines are BHAC15 isochrones for ages 0.5, 1, 2, 5, 10, 30, 100 and 1000 Myr.}

    \label{fig:hrd}
\end{figure*}

To evaluate whether the Module B SED-selected objects are consistent with being genuine young members of Wd2, we compare their location in the Hertzsprung-Russell diagram (HRD) with those found in the two comparison fields (i.e., Module A and CF). $T_\mathrm{eff}$ are taken from the adopted SED fits, and absolute F150W magnitudes are derived from the de-reddened photometry assuming they are located at the cluster distance. Fig.~\ref{fig:hrd} presents the resulting diagrams for the cluster, Module~A, and CF. The vast majority of Wd2 candidates lie above the 10~Myr isochrone, consistent with the young age of Wd2. In contrast, sources in the CF predominantly fall below this isochrone, occupying the region expected for background field populations. The candidates in Module~A span a broader range, with objects both above and below the 10~Myr line. This comparison indicates that the CF provides a good empirical representation of the contaminant population, while Module~A likely contains a mixture of contaminants and genuinely young, embedded cluster members. Motivated by this behavior, we divide the Wd2 candidates into two categories: Sources located above the 10~Myr isochrone are considered strong candidates (301 candidates), whereas those below it are flagged as weak candidates (52 candidates). Propagating the acceptable $T_\mathrm{eff}$--$A_V$ solutions into the HRD shows that only a small fraction of borderline objects change their classification relative to the adopted 10 Myr boundary. The clear separation between the Wd2 and CF populations remains unchanged, indicating that the interpretation of the HRD is not driven by uncertainties in the SED fitting. Sources with disk-related infrared excess can appear underluminous due to their disk-viewing geometry, we therefore kept those sources within the strong candidate sample regardless of their HRD position. All substellar candidates, estimated photospheric properties, and photometry are provided in Table~\ref{tab:candidate_parameters}.

A fraction of the substellar candidates occupy positions in the HR diagram above the youngest adopted isochrones. Moderate over-luminosity in young stellar objects may arise from several effects, including unresolved multiplicity, photometric variability, accretion, uncertainties in the evolutionary models, and uncertainties in the adopted cluster parameters such as distance and age. In particular, adopting a smaller distance for Wd2 would systematically reduce the inferred luminosities of all candidates. Nevertheless, 44 candidates remain more than 1 mag above the youngest isochrone, where these effects become increasingly difficult to reconcile with the observed luminosities. Only four of these candidates exhibit infrared excess indicative of circumstellar disks, a smaller proportion than in the remainder of the candidate sample. However, 11 of the 44 over luminous candidates (25\%) lack long-wavelength photometry beyond F200W, compared to $\sim$15\% in the rest of the sample, preventing an assessment of infrared excess for these objects and potentially contributing to this difference. Importantly, the SED-fitting behavior of these overluminous candidates does not differ from that of the remaining sample: their $\chi^2$ maps exhibit the same qualitative topology as the rest of the Wd2 candidates. Furthermore, these candidates are preferentially concentrated toward the central regions of Wd2 and are absent from the CF, both of which are consistent with cluster membership. While the origin of their over-luminosity therefore remains uncertain, we find no evidence from the SED fitting itself that these objects constitute a distinct population or are preferentially misclassified. Ultimately, spectroscopic follow-up will be required to determine the origin of their over-luminosity and establish their nature.

\begin{figure}[hbt!]
    \centering
    \includegraphics[width=\textwidth/21*10]{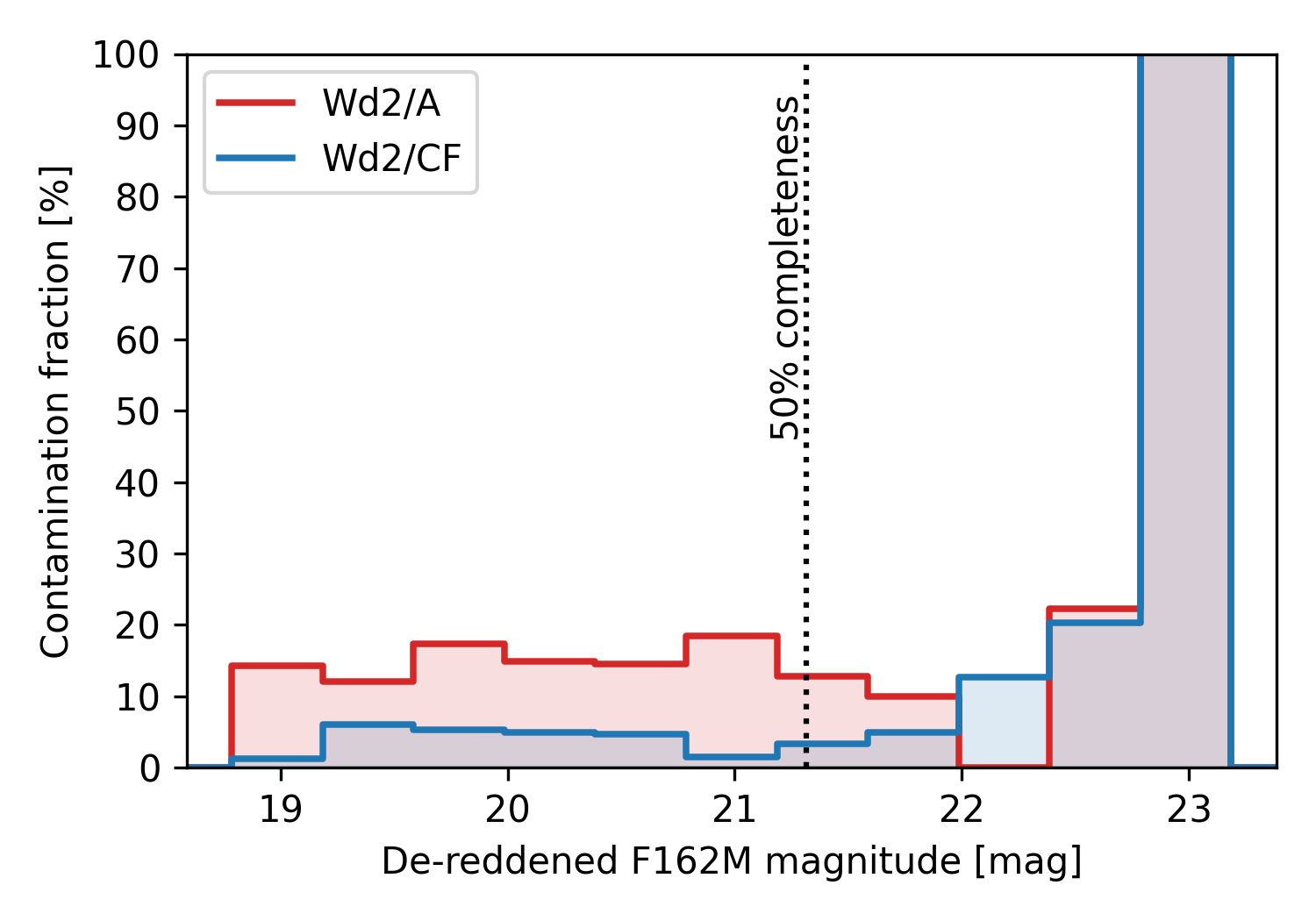}
    \caption{Distribution of the contamination fraction as a function of de-reddened F162M magnitude. Red line represents the number of strong candidates in Wd2 over those in Module A (red line), the blue line those of Wd2 over those in the CF (blue line). The vertical black dotted line represents the F162M 50\% completeness of the cluster assuming $A_V$=6.5 mag.}
    \label{fig:sed_cands_cont}
\end{figure}

We then estimate the contamination fraction of the strong candidate sample as a function of magnitude by comparing its number counts with those of sources that satisfy the same HRD selection (i.e., located above the 10 Myr isochrone) in the comparison fields. To account for different extinction between fields, this analysis is performed using de-reddened magnitudes. Importantly, both the Wd2 and comparison fields counts are corrected for photometric completeness described in Sect.~\ref{phot_completeness}. Specifically, each source is weighted by the inverse of the completeness at its F162M magnitude, as it is the filter that sets the effective completeness of our catalog. When comparing the Module B and CF, we correct the CF counts for the fact that the CF observations cover twice the sky area of the Module B field. Fig.~\ref{fig:sed_cands_cont} shows the resulting contamination fraction as a function of the F162M magnitude. The vertical dotted black line marks the de-reddened F162M magnitude corresponding to the 50\% completeness level in Module~B at the cluster extinction $A_V=6.5$ mag. We find that the contamination level inferred from the CF is $\sim$4\% down to F162M$\approx$22 mag. This magnitude is deeper than the 50\% completeness of the Module B and translates to a mass of $\sim$0.01-0.015~$M_{\odot}$ (10-15~$M_{\mathrm{Jup}}$), depending on the adopted cluster age. Performing the same comparison with Module~A yields somewhat higher contamination values, around $\sim$13\%. However, given the HRD evidence that Module~A itself likely contains genuine young substellar objects, this comparison is probably overestimating the contaminant contribution. At fainter magnitudes, the inferred contamination fraction rises rapidly. However, in this regime the results become increasingly uncertain due to the steep decline in detection efficiency and the larger completeness corrections applied. Overall, these results indicate that the majority of our strong SED-selected candidates in Module~B are consistent with being young members of Wd2, with low expected contamination down to $\sim$0.01-0.015~$M_{\mathrm{\odot}}$.

\subsection{Properties of identified substellar candidate sample}

\begin{figure}[hbt!]
    \centering
    \includegraphics[width=\textwidth/2]{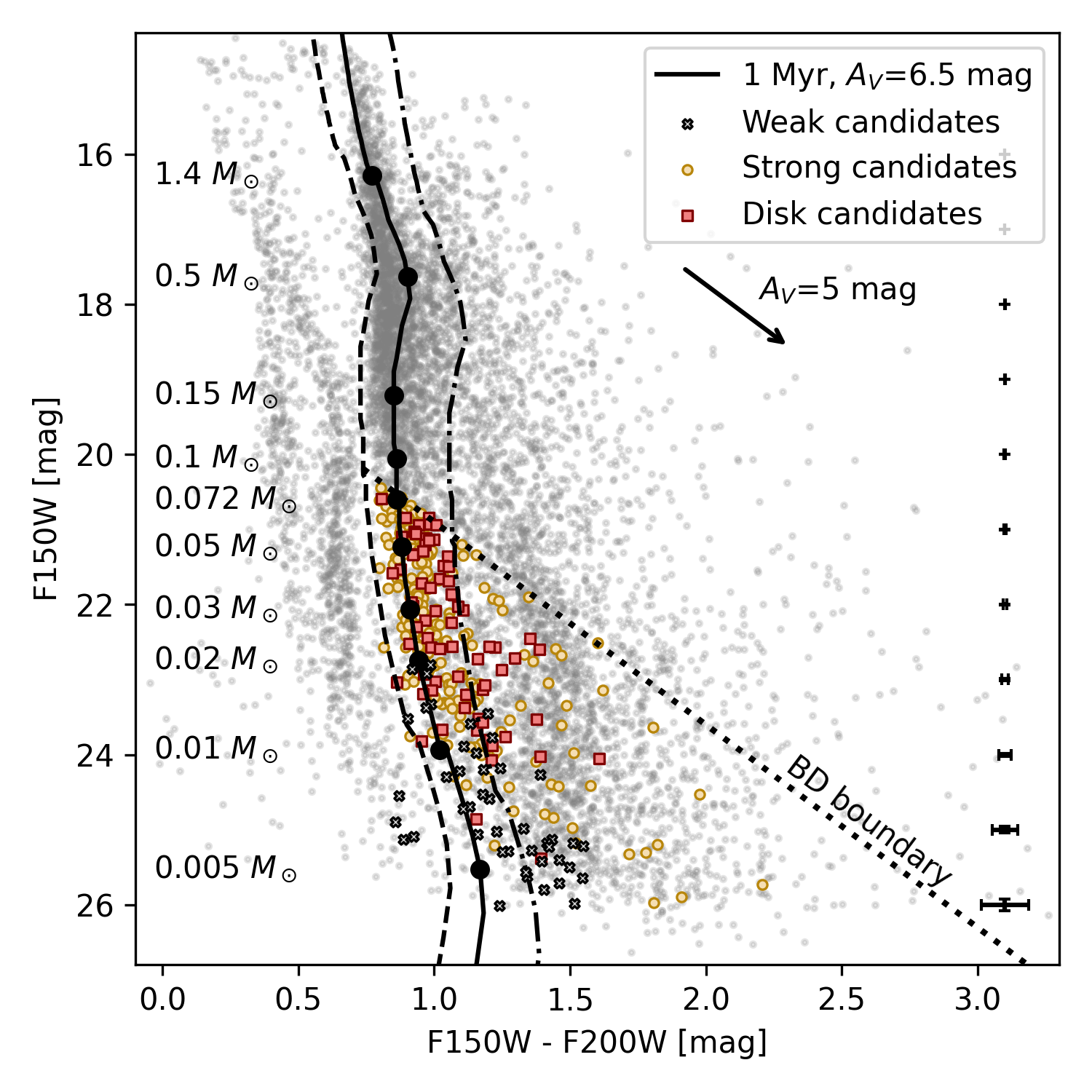}
    \caption{F150W$-$F200W vs. F150W CMD of the NIRCam observations of Wd2 (gray). The solid black line represents the stitched 1 Myr isochrone at a distance of 4.44 kpc and $A_V$=6.5 mag. Black dashed and dash-dotted lines represent the same isochrone with $A_V$=5 and 9 mag, respectively. We show the Wd2 strong (orange circle), infrared excess (red squares) and weak (black crosses) substellar candidates. Black errorbars at the right part of the plot represent the mean errors at different magnitudes.}
    \label{fig:cmd_wd2_cands}
\end{figure}

We show the position of the final Wd2 substellar candidates in the CMD in Fig.~\ref{fig:cmd_wd2_cands}. Strong candidates are plotted as orange circles, sources with infrared excess as red squares, and weak candidates as black crosses. The majority of the strong candidates lie close to the 1~Myr isochrone, in agreement with their HRD positions and consistent with membership to Wd2. Most of the weak candidates populate the faint end of the diagram, where we estimated that contamination from background reddened dwarfs may be present (see Fig.~\ref{fig:app_cmd_reddened}). We also identify a subset of strong candidates located redward of the $A_V = 9$ mag isochrone (black dash-dotted line). The majority of these red sources are located over areas of the cluster with increased local extinction (B2 and B4 detectors located in the bottom panels of Module B in Fig.~\ref{fig:rgb_wd2}), likely explaining their redder colors.

\begin{figure}[hbt!]
    \centering
    \includegraphics[width=\textwidth/2]{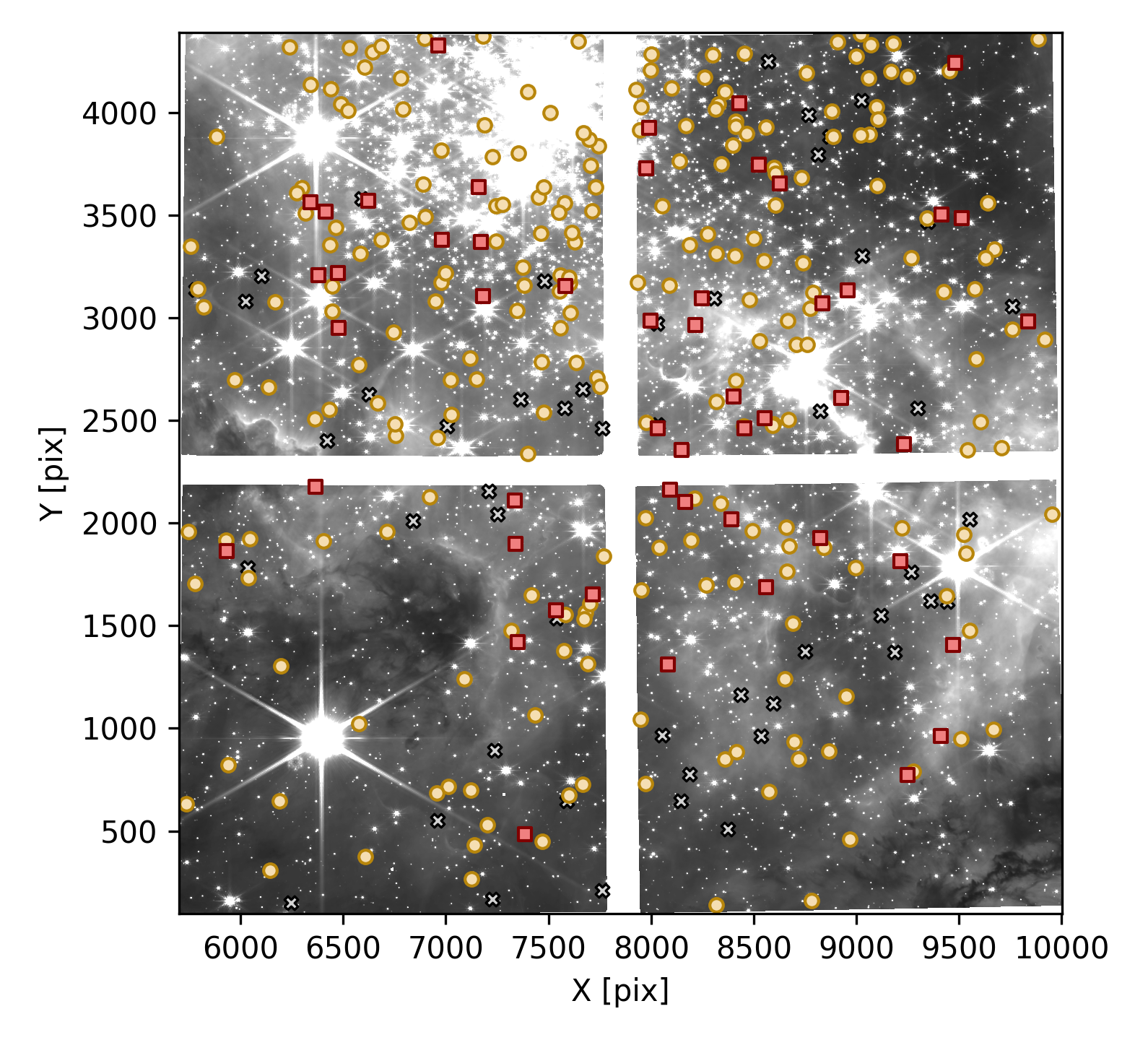}
    \caption{Spatial distribution of the strong (orange circles), weak (black crosses) and disk-bearing candidates (red squares). Background image is the F200W mosaic.}
    \label{fig:spatial}
\end{figure}

Fig.~\ref{fig:spatial} shows the spatial distribution of the identified substellar candidates across Module~B. The strong and disk-bearing candidates are preferentially concentrated toward the central regions of Wd2, with most sources located in the vicinity of the main cluster clump. Relatively few candidates are recovered within the innermost region of the primary cluster core, as expected from the artificial-star tests presented in Sect.~\ref{phot_completeness}, which show a strong reduction in sensitivity within the central $25''$ due to crowding and bright saturated stars. Candidates are also found throughout the surrounding field, including regions of enhanced nebulosity and extinction. Overall, their projected distribution is qualitatively consistent with an association with Wd2. In contrast, the weak candidates do not exhibit a similarly concentrated spatial distribution, consistent with the higher level of contamination expected for that sample.

\subsection{Disk-bearing population}

In Sect.~\ref{results_sed_wd2}, we identified 73 substellar candidates presenting infrared excess consistent with the presence of disks based on visual inspection of their SEDs. Among the strong candidates, disk-bearing sources account for $21.6^{+2.6}_{-2.4}$\% of the sample, and $27.7^{+3.4}_{-3.2}$\% when considering only objects detected in F410M.

\begin{figure}[hbt!]
    \centering
    \includegraphics[width=\textwidth/21*10]{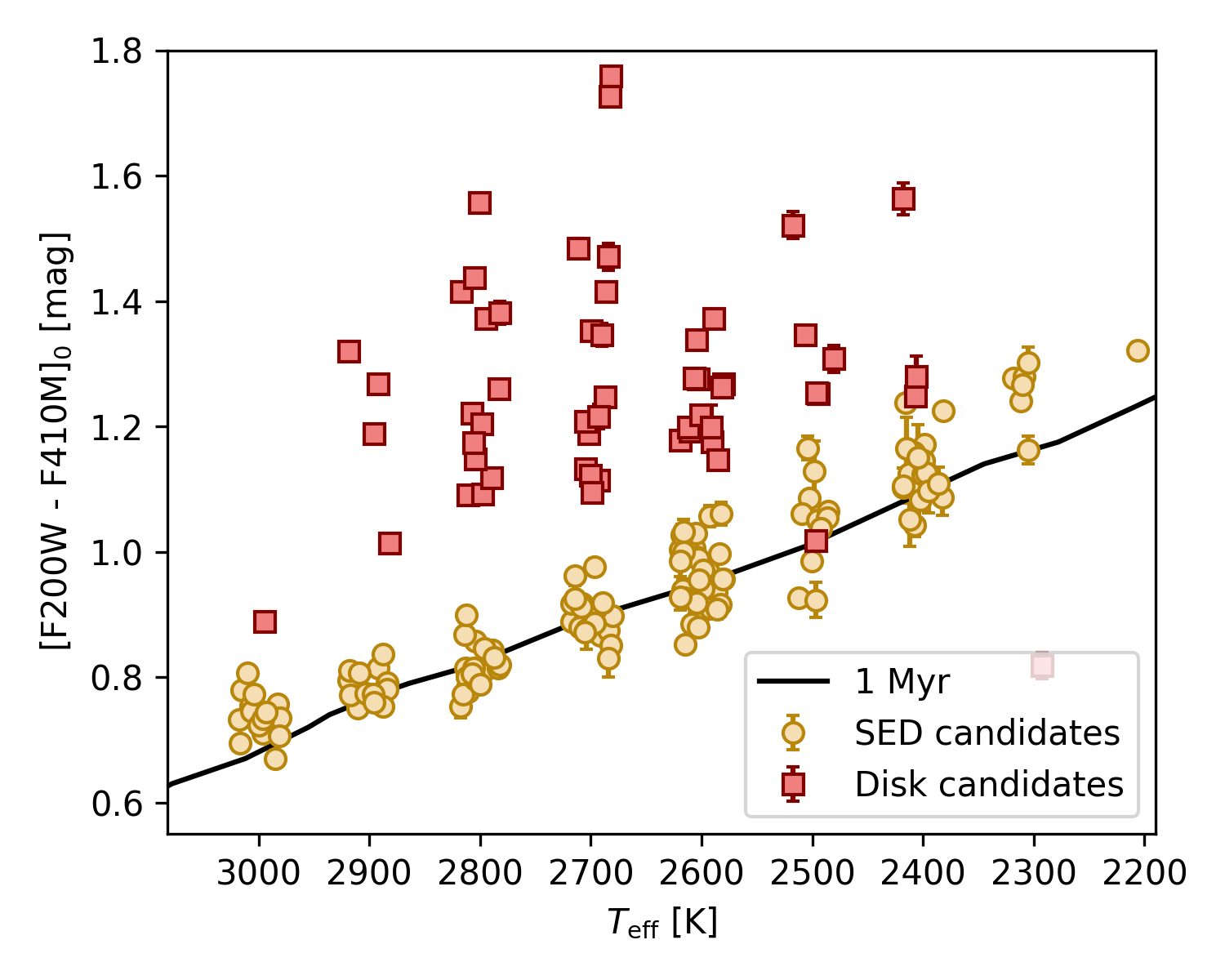}
    \caption{De-reddened F200W-F410M color as a function of the best fit SED $T_\mathrm{eff}$ for strong SED (orange circles) and disk-bearing candidates (red squares). A random $\pm$20 K offset was applied to all sources for clarity. The black solid line represents the stitched 1 Myr isochrone.}
    \label{fig:ir_excess_f200w_f410m}
\end{figure}

A common approach to identifying disk-bearing sources relies on de-reddened colors that combine near- and mid-infrared photometry from WISE or Spitzer, where disk-bearing objects appear redder than expected from naked photospheres. Within the wavelength coverage of our NIRCam observations, the F200W$-$F410M color provides a diagnostic broadly analogous to the commonly used $K$--IRAC2 color. Fig.~\ref{fig:ir_excess_f200w_f410m} shows the de-reddened F200W$-$F410M color as a function of the best-fit $T_\mathrm{eff}$ for the Wd2 substellar candidates. Candidates without evidence of infrared excess generally follow the photospheric locus traced by the cluster isochrone, whereas the visually identified disk-bearing candidates occupy systematically redder colors. The separation is comparable to that observed in $K$--IRAC colors of young BDs \citep{almendros23}, supporting the interpretation that the long-wavelength excesses identified from the SEDs arise from circumstellar emission. Two candidates identified in Sect.~\ref{results_sed_wd2} as having infrared excess fall well-below the disk-bearing locus. Both, however, exhibit clear excess at the other $>2\,\mu$m filters, indicating that this discrepancy is likely due to issues with the F410M photometry. In one case, the source lies along the diffraction spikes of a nearby saturated star, rendering its F410M measurement unreliable. 

If the present candidate sample were representative of the full Wd2 substellar population, the observed disk-bearing fraction would be lower than the substellar disk fractions commonly reported in nearby 1--2 Myr star-forming regions such as Taurus, Chamaeleon~I, and NGC~1333 \citep[40--60\%; e.g.,][]{luhman10_tau,luhman16_perseus}. However, this comparison should be treated cautiously, as the available wavelength coverage may miss evolved or transition disks whose excess becomes detectable only beyond the NIRCam range. Although JWST/MIRI observations of Wd2 are available from the same program, its sensitivity does not reach the substellar regime (Monsch et al., in prep.).

\section{Conclusions}
\label{summary}

We have presented deep JWST/NIRCam imaging of the supermassive young cluster Wd2, obtained as part of the EWOCS program. The presented dataset comprises 10 wide and medium-band filters spanning 1.15--4.1~$\mu$m, designed to enable a comprehensive analysis of the cluster. The observations reach a 50\% completeness limit at 0.015--0.02~$M_\odot$ ($\sim$15--20~$M_\mathrm{Jup}$) at the adopted distance (4.44 kpc), age (1--2 Myr), and extinction ($A_V=6.5$ mag) of the cluster, providing the deepest view of the substellar population of a supermassive star cluster to date. PSF photometry was performed using DOLPHOT, with sub-pixel astrometric alignment across filters.

The main focus of this work is the identification and characterization of candidate substellar members through the comparison of their SEDs with atmospheric models. Our main findings are summarized as follows:

\begin{itemize}

\item We validated the SED fitting methodology using synthetic JWST photometry of spectroscopically confirmed young brown dwarfs together with simulated field contaminants. The methodology reliably recovers known young brown dwarfs while remaining robust against contamination from reddened evolved stars. The only stellar contaminants naturally satisfying the adopted temperature criterion are intrinsically cool field dwarfs, whose contribution is expected to be limited by their restricted location in CMD space.

\item Applying the validated methodology to the Wd2 observations, we identify 353 substellar candidates. Based on their position in the HR diagram, we classify 301 as strong candidates and 52 as weak candidates.

\item Comparison with the CF indicates that the contamination level of the strong candidate sample is $\lesssim5$\% down to masses of $\sim$0.01--0.015~$M_\odot$, below the 50\% completeness limit of the observations.

\item The strong candidates follow the expected cluster sequence in the CMD and are preferentially concentrated toward the central regions of Wd2, consistent with cluster membership. In contrast, the weak candidates are predominantly located in the region of the HR diagram and CMD where contamination from background field dwarfs is expected to be higher, and they exhibit a less centrally concentrated spatial distribution, consistent with their higher expected contamination.

\item We identify 73 substellar candidates whose SEDs exhibit infrared excess consistent with circumstellar disks. These candidates occupy the expected infrared-excess locus in the de-reddened F200W$-$F410M versus $T_\mathrm{eff}$ diagram, supporting the interpretation that the observed infrared excesses arise from circumstellar emission. Among the candidates with reliable F410M photometry, they represent $27.7^{+3.4}_{-3.2}$\% of the sample. If the present candidate sample is representative of the underlying substellar population, this fraction would be lower than those typically reported in nearby low-density star-forming regions, although this comparison remains tentative because the observations do not cover wavelengths beyond the NIRCam range, where more evolved disks may become detectable.

\end{itemize}

Overall, this work demonstrates the capability of JWST/NIRCam medium- and wide-band photometry to identify robust samples of candidate BDs in distant, embedded, and crowded massive star clusters using imaging observations alone. The resulting candidate catalog provides an ideal basis for future spectroscopic studies of the substellar population of Wd2.

\begin{acknowledgements}

V.A-A and M.G.G acknowledge support from the INAF grant 1.05.12.05.03 and 1.05.24.07.02. K.M. acknowledges support from the Fundação para a Ciência e a Tecnologia (FCT) through the CEEC-individual contract 2022.03809.CEECIND and grant UID/04434/2023, and the Scientific Visitor Programme of the European Southern Observatory (ESO) in Chile. A.S acknowledges support from the UKRI Science and Technology Facilities Council through grant ST/Y001419/1/. A.B acknowledges support from the Deutsche Forschungsgemeinschaft (DFG, German Research Foundation) under Germany's Excellence Strategy - EXC 2094 - 390783311. A.G. acknowledges support from the NSF under CAREER 2142300. K.M. was supported by JWST-GO-1905 and JWST-GO-3523. E.S. is supported by the international Gemini Observatory, a program of NSF NOIRLab, which is managed by the Association of Universities for Research in Astronomy (AURA) under a cooperative agreement with the U.S. National Science Foundation, on behalf of the Gemini partnership of Argentina, Brazil, Canada, Chile, the Republic of Korea, and the United States of America. R.B. acknowledge support the INAF grant 1.05.23.04.02. T.J.H  acknowledges a Dorothy Hodgkin Fellowship, UKRI guaranteed funding for a Horizon Europe ERC consolidator grant (EP/Y024710/1) and and UKRI/STFC grant ST/X000931/1.

This work is based on observations made with the NASA/ESA/CSA James Webb Space Telescope. The data were obtained from the Mikulski Archive for Space Telescopes at the Space Telescope Science Institute, which is operated by the Association of Universities for Research in Astronomy, Inc., under NASA contract NAS 5-03127 for JWST. These observations are associated with program \#3523.

\end{acknowledgements}

\bibliographystyle{aa} 
\bibliography{wd2_bds.bib}

\begin{appendix}

\section{$\chi^2$ maps of validation and candidates targets}
\label{results_sed_validation}

\begin{figure*}[hbt!]
    \centering
    \includegraphics[width=\textwidth]{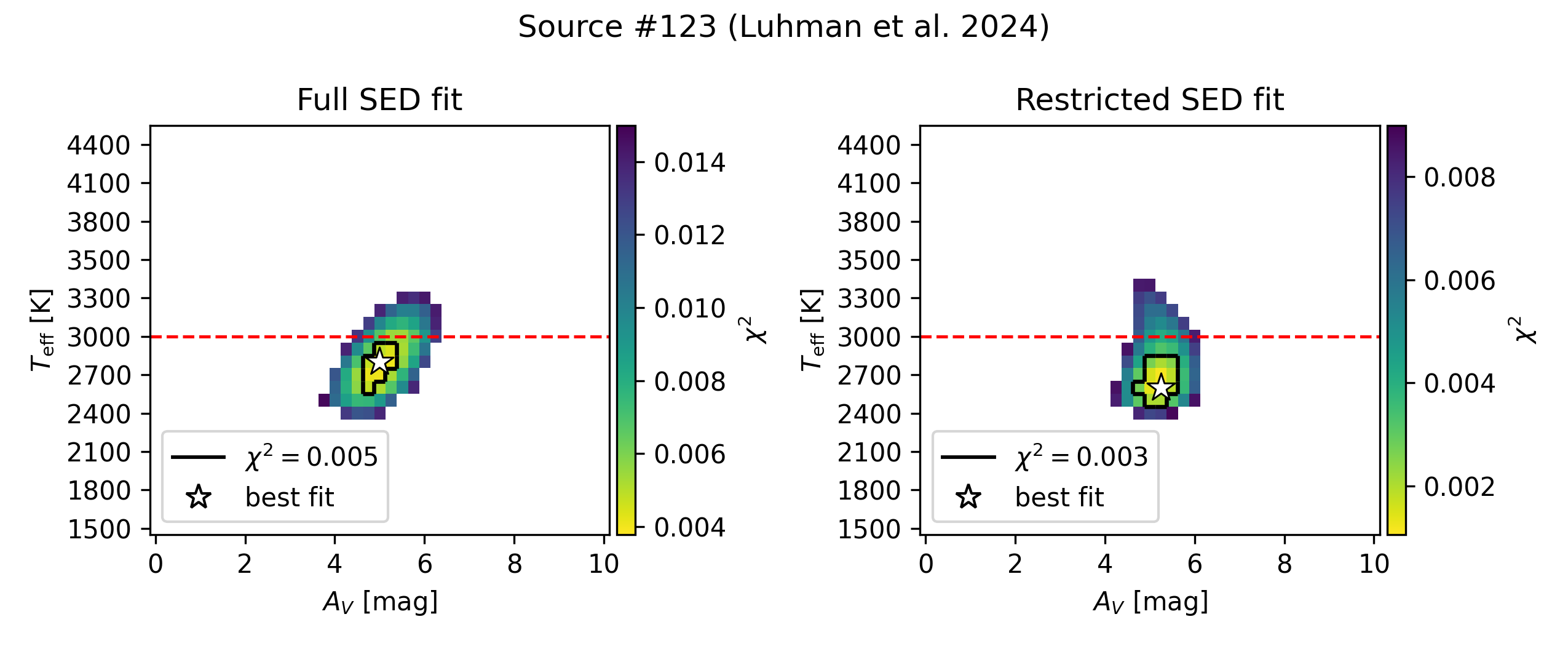}
    \includegraphics[width=\textwidth]{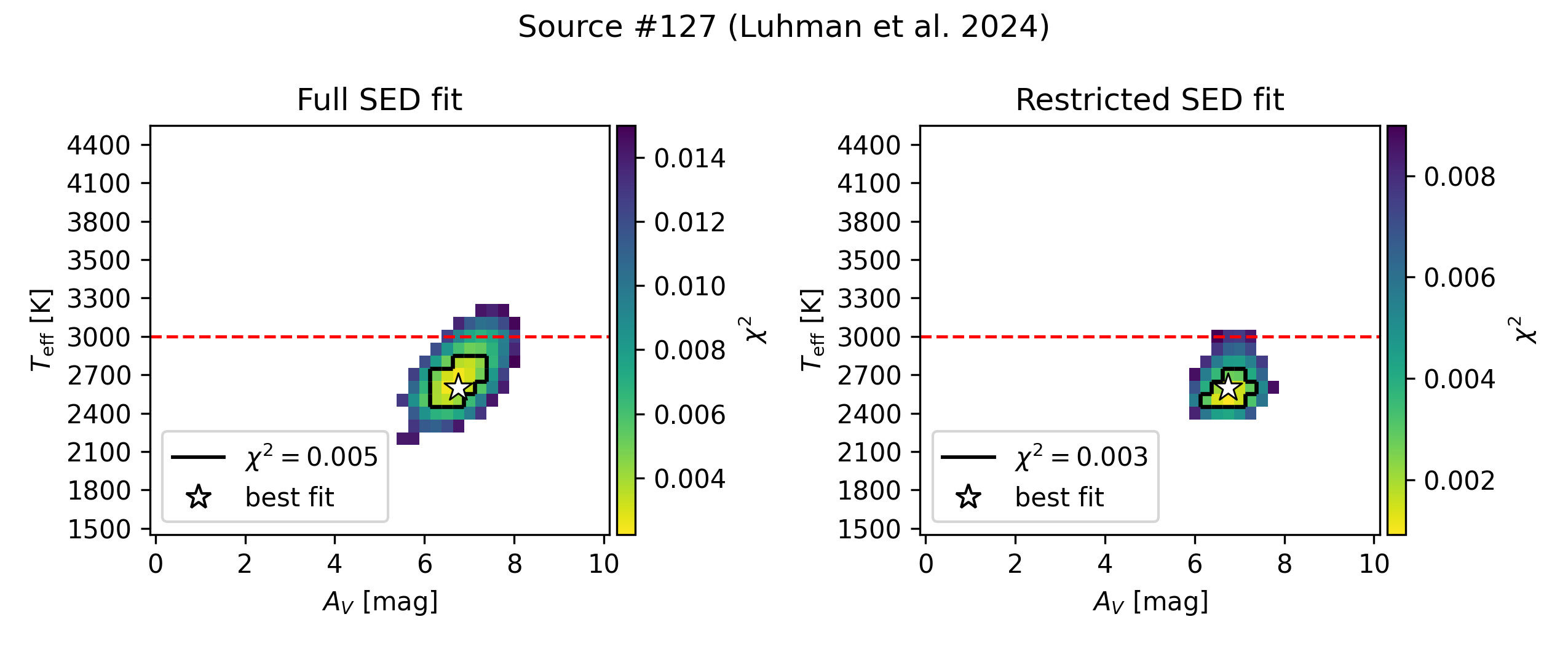}
    \includegraphics[width=\textwidth]{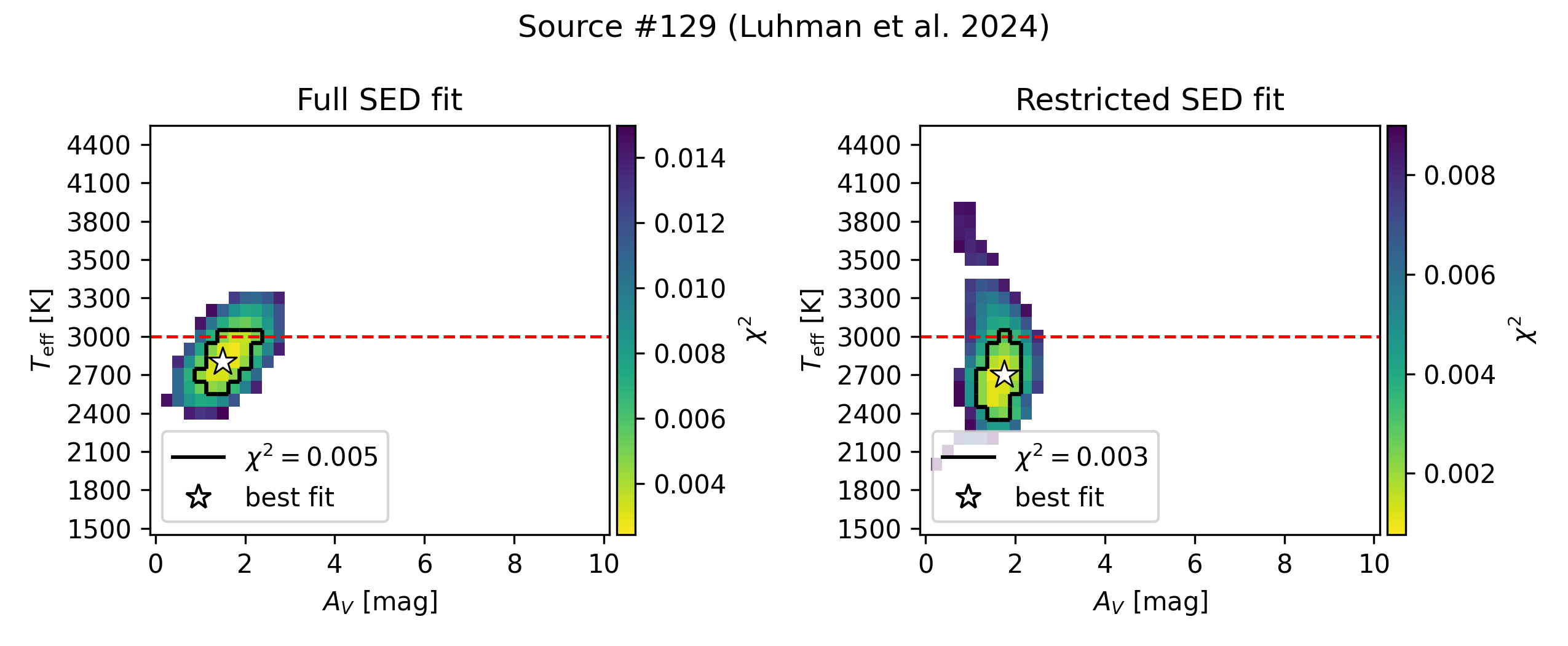}
    \caption{Example $\chi^2$ maps in the $T_\mathrm{eff}$--$A_V$ plane for three bona fide young brown dwarfs from the validation set. The left panel shows the results obtained using the full SED, while the right panel corresponds to the restricted fit. The color scale shows the $\chi^2$ value, limited to three times the adopted acceptance threshold, while the black contour encloses the grid points satisfying the adopted $\chi^2$ criterion, and the white star marks the best-fitting model. The red dashed line indicates the adopted stellar/substellar boundary at $T_\mathrm{eff}=3000$~K.}
    \label{fig:chi2_map}
\end{figure*}

\begin{figure*}[hbt!]
    \centering
    \includegraphics[width=\textwidth]{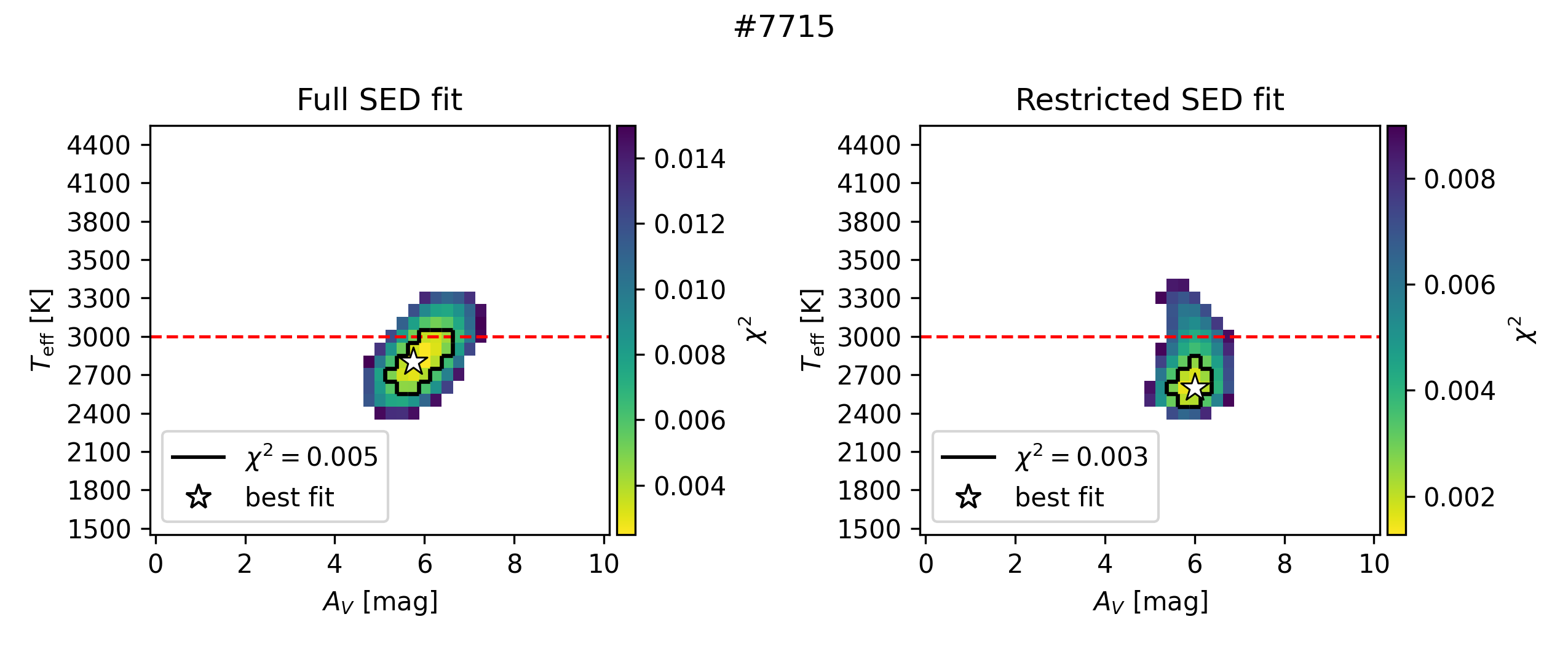}
    \includegraphics[width=\textwidth]{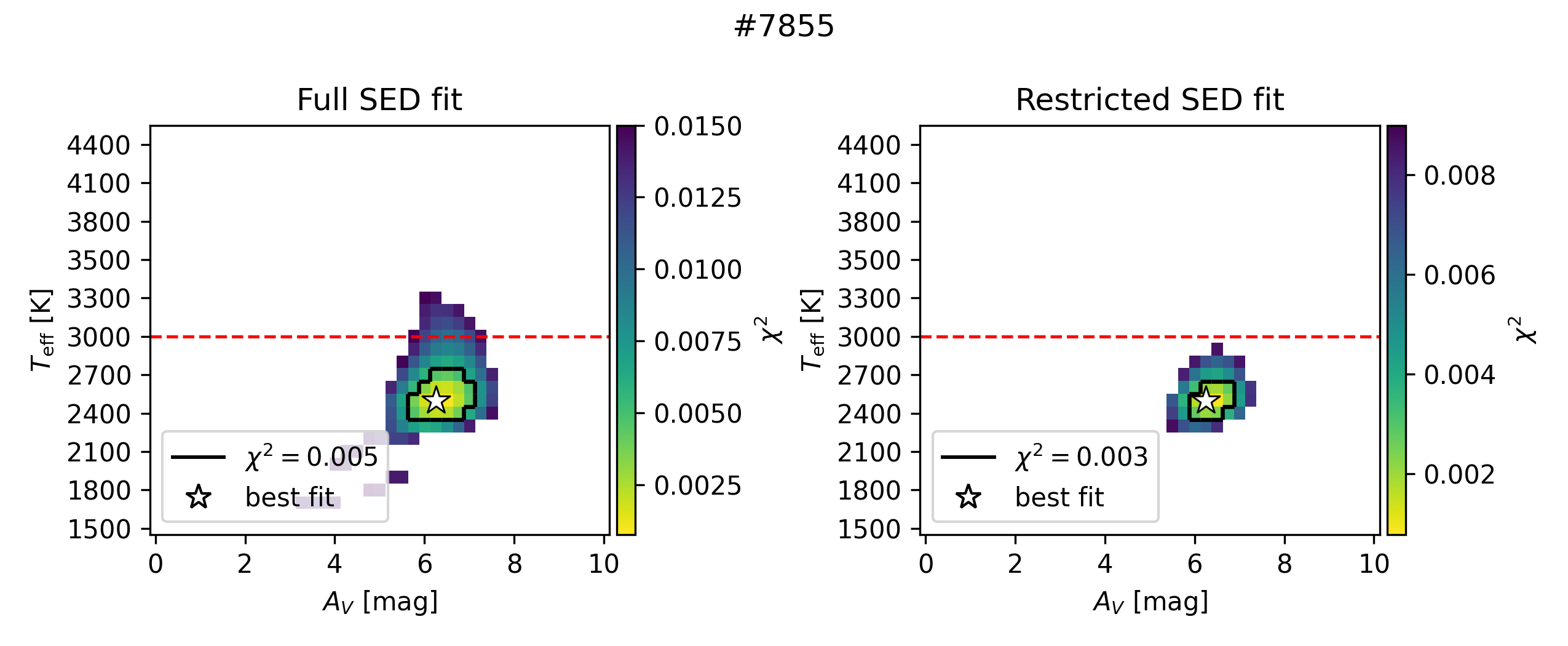}
    \includegraphics[width=\textwidth]{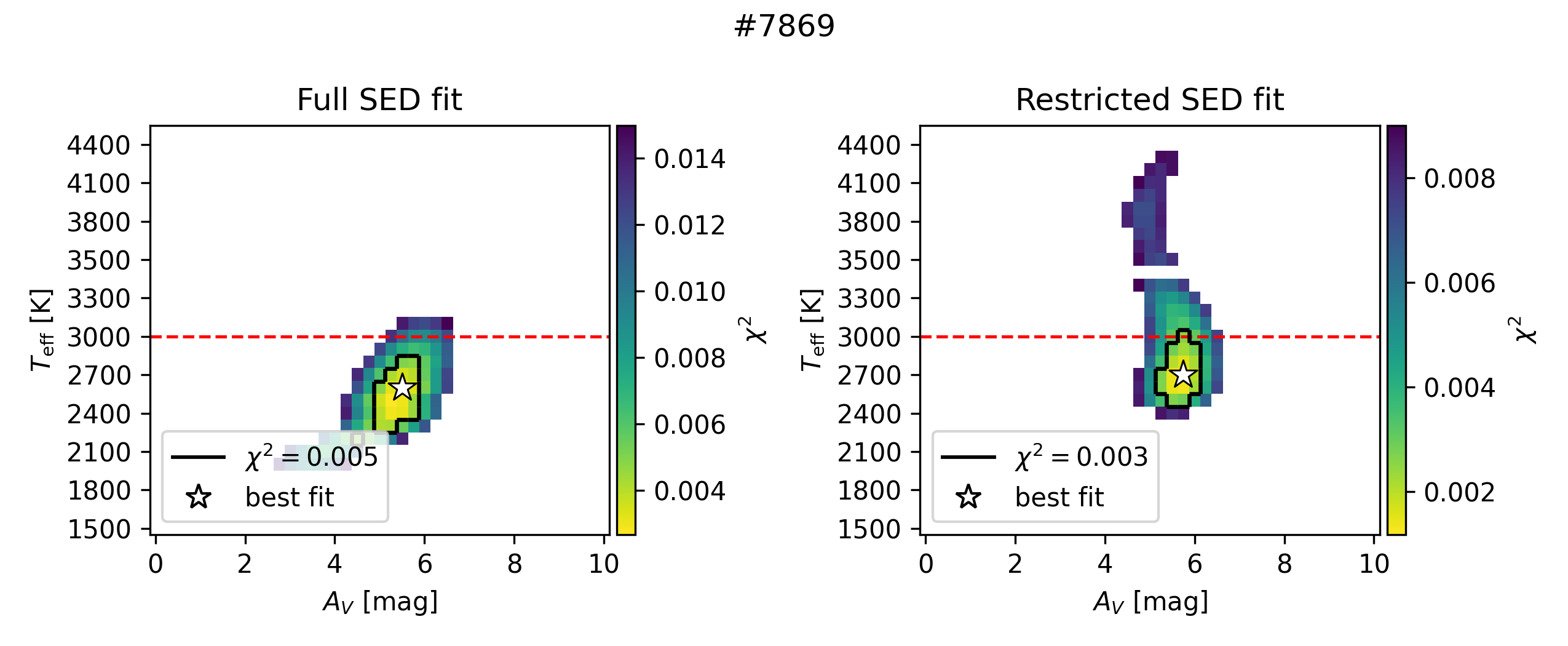}
    \caption{Example $\chi^2$ maps in the $T_\mathrm{eff}$--$A_V$ plane for three Wd2 substellar candidates. Figure layout is the same as Fig.~\ref{fig:chi2_map}.}
    \label{fig:chi2_map_cands}
\end{figure*}

Figs.~\ref{fig:chi2_map} and \ref{fig:chi2_map_cands} show full $\chi^2$ surface over the $T_\mathrm{eff}$--$A_V$ grid for three validation and substellar candidate targets, respectively.

\section{Properties of identified candidates}

In Table~\ref{tab:candidate_parameters} we present the substellar candidates identified in Sect.~\ref{results_sed}, their estimated properties and photometry.

\begin{sidewaystable*}
\caption{ID, position, estimated photospheric parameters, presence of disk, candidate type (strong: S, weak: W) and photometry of all substellar candidates.}
\label{tab:candidate_parameters}
\centering
\begin{tabular}{l c c c c c c c c c c c c c c c c}
\hline\hline
ID & RA & Dec & $T_{\rm eff}$ & $A_V$ & Disk & Type & F115W & F150W & F162M & F182M & F200W & F250M & F277M & F300M & F335M & F410M \\
\hline
7383 & 156.008358 & -57.755895 & $2900^{+100}_{-400}$ & $5.25^{+0.50}_{-0.75}$ & N & S & 21.63 & 20.54 & 20.03 & 19.87 & 19.71 & 19.27 & 19.16 & 19.19 & 18.84 & 18.39 \\
7385 & 156.011207 & -57.745266 & $2800^{+300}_{-300}$ & $5.00^{+0.75}_{-0.25}$ & N & S & 21.60 & 20.44 & 19.96 & 19.84 & 19.64 & 19.28 & 19.20 & 19.11 & 18.74 & 18.36 \\
7425 & 156.016077 & -57.757456 & $2900^{+200}_{-400}$ & $5.25^{+0.25}_{-1.00}$ & N & S & 21.65 & 20.60 & 20.10 & 19.94 & 19.77 & 19.34 & 19.29 & 19.20 & 18.89 & 18.49 \\
7488 & 156.005114 & -57.744709 & $3000^{+300}_{-500}$ & $5.50^{+0.50}_{-0.75}$ & N & S & 21.73 & 20.60 & 20.11 & 19.95 & 19.75 & 19.38 & 19.32 & 19.16 & 18.90 & 18.52 \\
7491 & 156.027214 & -57.742132 & $2900^{+100}_{-500}$ & $5.50^{+0.75}_{-0.50}$ & N & S & 21.73 & 20.58 & 20.05 & 19.90 & 19.70 & 19.36 & 19.29 & 19.20 & 18.82 & 18.39 \\
7507 & 156.000574 & -57.754610 & $3000^{+300}_{-500}$ & $5.75^{+0.50}_{-0.75}$ & N & S & 21.79 & 20.60 & 20.11 & 19.93 & 19.76 & 19.34 & 19.30 & 19.20 & 18.90 & 18.46 \\
7527 & 156.010596 & -57.741353 & $2700^{+1200}_{-100}$ & $5.25^{+0.50}_{-0.75}$ & Y & S & 21.79 & 20.59 & 20.17 & 19.97 & 19.79 & 19.39 & 19.29 & 19.18 & 18.86 & 18.38 \\
7537 & 156.007094 & -57.758243 & $2400^{+200}_{-400}$ & $5.25^{+0.50}_{-2.00}$ & N & S & 21.83 & 20.79 & 20.20 & 20.12 & 19.87 & -- & -- & -- & -- & -- \\
7548 & 156.021761 & -57.755031 & $3000^{+1700}_{-400}$ & $6.25^{+0.50}_{-1.50}$ & N & S & 21.82 & 20.57 & 20.12 & 19.90 & 19.69 & 19.27 & 19.13 & 19.01 & 18.90 & 18.32 \\
7549 & 156.029714 & -57.750571 & $2900^{+100}_{-500}$ & $5.25^{+0.75}_{-0.50}$ & N & S & 21.82 & 20.68 & 20.17 & 20.02 & 19.83 & 19.48 & 19.42 & 19.35 & 18.95 & 18.56 \\
\hline
\end{tabular}
\tablefoot{The complete table is available at the CDS.}
\end{sidewaystable*}

\end{appendix}

\end{document}